\documentstyle[twocolumn,prd,aps,epsf]{revtex}
\begin{document}
\draft
%
%
\newcommand{\nc}{\newcommand}
\nc{\bea}{\begin{eqnarray}}
\nc{\eea}{\end{eqnarray}}
\nc{\beq}{\begin{equation}}
\nc{\eeq}{\end{equation}}
\nc{\bi}{\begin{itemize}}
\nc{\ei}{\end{itemize}}
\nc{\la}[1]{\label{#1}}
\nc{\doo}{\partial}
\nc{\half}{\frac{1}{2}}
\nc{\third}{\frac{1}{3}}
\nc{\quarter}{\frac{1}{4}}
\nc{\EHe}{{}^3{\rm He}}
\nc{\UHe}{{}^4{\rm He}}
\nc{\ZLi}{{}^7{\rm Li}}
\nc{\DH}{{\rm D}/{\rm H}}
\nc{\ZLiH}{{}^7{\rm Li}/{\rm H}}
\nc{\et}{\eta_{10}}
\nc{\etl}{\eta_{\rm low}}
\nc{\eth}{\eta_{\rm high}}
\nc{\GeV}{\mbox{ GeV}}
\nc{\MeV}{\mbox{ MeV}}
\nc{\keV}{\mbox{ keV}}
\nc{\req}{\hat r_{\rm eq}}
\nc{\ropt}{r_{\rm opt}}
\nc{\cc}{c.c.}
\nc{\ssh}{s.s.}
\nc{\nhigh}{n_{\rm high}}
\nc{\nlow}{n_{\rm low}}
\nc{\nmean}{n_{\rm mean}}
\nc{\dhigh}{D_{\rm high}}
\nc{\diflen}{\sqrt{Dt}}
\nc{\diflenns}{\sqrt{Dt_{\rm ns}}}
\nc{\etal}{{\it et al.}}
\nc{\lsim}{\mbox{\raisebox{-.6ex}{~$\stackrel{<}{\sim}$~}}}
\nc{\gsim}{\mbox{\raisebox{-.6ex}{~$\stackrel{>}{\sim}$~}}}
\nc{\x}[1]{}
%
%

\nc{\AJ}[3]{{Astron.~J.\ }{{\bf #1}{, #2}{ (#3)}}}
\nc{\anap}[3]{{Astron.\ Astrophys.\ }{{\bf #1}{, #2}{ (#3)}}}
\nc{\ApJ}[3]{{Ap.~J.\ }{{\bf #1}{, #2}{ (#3)}}}
\nc{\apjl}[3]{{Ap.~J.\ Lett.\ }{{\bf #1}{, #2}{ (#3)}}}
\nc{\app}[3]{{Astropart.\ Phys.\ }{{\bf #1}{, #2}{ (#3)}}}
\nc{\araa}[3]{{Ann.\ Rev.\ Astron.\ Astrophys.\ }{{\bf #1}{, #2}{ (#3)}}}
\nc{\arns}[3]{{Ann.\ Rev.\ Nucl.\ Sci.\ }{{\bf #1}{, #2}{ (#3)}}}
\nc{\arnps}[3]{{Ann.\ Rev.\ Nucl.\ and Part.\ Sci.\ }{{\bf #1}{, #2}{ (#3)}}}
\nc{\mpl}[3]{{Mod.\ Phys.\ Lett.\ }{{\bf #1}{, #2}{ (#3)}}}
\nc{\Nat}[3]{{Nature }{{\bf #1}{, #2}{ (#3)}}}
\nc{\ncim}[3]{{Nuov.\ Cim.\ }{{\bf #1}{, #2}{ (#3)}}}
\nc{\nast}[3]{{New Astronomy }{{\bf #1}{, #2}{ (#3)}}}
\nc{\np}[3]{{Nucl.\ Phys.\ }{{\bf #1}{, #2}{ (#3)}}}
\nc{\pr}[3]{{Phys.\ Rev.\ }{{\bf #1}{, #2}{ (#3)}}}
\nc{\PRD}[3]{{Phys.\ Rev.\ D\ }{{\bf #1}{, #2}{ (#3)}}}
\nc{\PRL}[3]{{Phys.\ Rev.\ Lett.\ }{{\bf #1}{, #2}{ (#3)}}}
\nc{\PL}[3]{{Phys.\ Lett.\ }{{\bf #1}{, #2}{ (#3)}}}
\nc{\prep}[3]{{Phys.\ Rep.\ }{{\bf #1}{, #2}{ (#3)}}}
\nc{\RMP}[3]{{Rev.\ Mod.\ Phys.\ }{{\bf #1}{, #2}{ (#3)}}}
\nc{\rpp}[3]{{Rep.\ Prog.\ Phys.\ }{{\bf #1}{, #2}{ (#3)}}}
\nc{\ibid}[3]{{\sl ibid.\ }{{\bf #1}{, #2}{ (#3)}}}
%
%

\title{Inhomogeneous Big-Bang Nucleosynthesis in Light of
       Recent Observations}

\author{K.~Kainulainen$^{a,}$\cite{mailk},
        H.~Kurki-Suonio$^{b,}$\cite{mailh},
        and E.~Sihvola$^{c,}$\cite{maile}}

\address{
  $^a$NORDITA, Blegdamsvej 17, DK-2100, Copenhagen \O , Denmark \\
  $^b$Helsinki Institute of Physics,
      P.O.Box 9, FIN-00014 University of Helsinki, Finland \\
  $^c$Department of Physics,
      P.O.Box 9, FIN-00014 University of Helsinki, Finland}

\date{July 9, 1998}

\maketitle

\begin{abstract}
We consider inhomogeneous big bang nucleosynthesis in light of the present
observational situation.  Different observations of $\UHe$ and D disagree with
each other, and depending on which set of observations one uses, the estimated
primordial $\UHe$ corresponds to a lower baryon density in standard big bang
nucleosynthesis than what one gets from deuterium.  Recent Kamiokande results
rule out a favorite particle physics solution to this tension between $\UHe$
and D.  Inhomogeneous nucleosynthesis can alleviate this tension, but
the more likely solution is systematics in the observations.
The upper limit to $\Omega_b$ from inhomogeneous nucleosynthesis is higher
than in standard nucleosynthesis, given that the distance scale of the
inhomogeneity is near the optimal value, which maximizes effects of neutron
diffusion.  Possible sources of baryon inhomogeneity include the QCD and
electroweak phase transitions.  The distance scale of the inhomogeneities
arising from the electroweak transition is too small for them to have a
large effect on nucleosynthesis, but the effect may still be larger than
some of the other small corrections recently incorporated to SBBN codes.
\end{abstract}


\narrowtext

\vspace*{-14.0cm}
\noindent
\hspace*{11.0cm} \mbox{HIP-1998-38/TH,\quad NORDITA-98/47 HE}
\vspace*{13.0cm}

%
%
\section{Introduction}

Standard big bang nucleosynthesis \cite{SBBN,BS85,WSSOK,SKM93} (SBBN)
predicts the  primordial abundances of D, $\EHe$, $\UHe$, and $\ZLi$ as a
function of a single parameter, the baryon-to-photon ratio $\eta \equiv
n_b/n_\gamma$, which is related to the baryonic mass-density parameter
$\Omega_b \equiv 8\pi G\rho_{b0}/3H_0^2$ by
\beq
    \Omega_b h^2 = 3.70\times10^{-3}\et,
\eeq
where $\et \equiv 10^{10}\eta$ and $h\equiv H_0/100 {\rm km s}^{-1}
{\rm Mpc}^{-1}$. The observed abundances of these isotopes are in a rough
agreement with the SBBN predictions \cite{CST95} for a range of $\et$, which
is compatible with other cosmological bounds on the amount of baryonic matter
in the universe.
In principle, comparing SBBN predictions with primordial abundances
extrapolated from observations pins down
the precise value of $\et$. A few years ago the standard result was $\et
\sim$ 3--4 \cite{WSSOK,SKM93}, but even much tighter constraints were
published (e.g., $2.69 \leq \et \leq 3.12$ \cite{KK94}).  Recently the
situation has become more complicated, and it seems that such precise
determinations were premature.

Since the discovery of the $\tau$ lepton, implying three flavours of light
neutrinos, there has been tension between $\UHe$ and D
in SBBN \cite{tension,Dicus}.  Olive \etal \cite{OSS97} (OSS97) have
reviewed the $\UHe$
observations and their best estimate is
\beq
   Y_p = 0.230 \pm 0.003.
\eeq
This corresponds to $\et = 1.4 \pm 0.3$ and hence to primordial $\DH \sim
\hbox{2--3}\times10^{-4}$ in SBBN,
whereas the present D/H in the ISM is\cite{ISMobs}
only $1.5\times10^{-5}$. Most models of galactic chemical evolution have
difficulty explaining this much deuterium astration \cite{Dchemev}, and
prefer a much lower primordial D/H and thus a higher baryon density,
$\et \sim 5$.

The conventional way to deal with this tension has been to compromise by
settling on an intermediate $\et$ which is preferred neither by $\UHe$ nor
by D/H but is considered acceptable to both.  This however, leads to an
artificially high precision in the $\et$ determination, because while the
individual ranges in $\et$ accepted by $\UHe$ and D/H are wide,
their overlap is narrow. Tension increased when data was subjected
to more thorough formal statistical analysis, culminating in a claim of
a ``crisis'' in SBBN, by Steigman \etal \cite{crisis}, who concluded that
given the existing data the overlap is in fact nonexistent.

In the context of SBBN the resolution of this crisis requires either a
revision of the picture of the galactic chemical evolution \cite{fkot},
so that much more deuterium astration can be accommodated \cite{chemevunc},
or a large systematic error in the $Y_p$ determination from the observations
\cite{CST95,KS96,He4unc}. Indeed,
based on a number of new $\UHe$ observations,
Izotov and Thuan \cite{IT98}
have claimed a significantly higher $Y_p$ than OSS97:
\beq
   Y_p = 0.244 \pm 0.002.
\eeq
Whether this new value is to be accepted as such over the old OSS97 value
is yet unclear, since several sources of poorly known systematic effects
are expected to contribute to the discrepancy \cite{trento}.

Interestingly, some particle physics solutions based on a massive decaying
tau neutrino \cite{kawa} can now be ruled out using the recent results from
Kamiokande \cite{kamioka}. The directional dependence in the upward going muon
neutrino deficiency seen in the Super Kamiokande experiment is a strong
implication that the muon neutrinos undergo oscillations while traversing
through the earth. This implies that $\nu_\mu$ mixes with either a tau
neutrino
or a new sterile neutrino with a mass splitting of about $\delta m^2 \sim
10^{-3} \mbox{eV}^2$ and with an almost maximal mixing angle.
If this mixing is between $\nu_\mu$
and $\nu_\tau$, then $\nu_\tau$ is obviously light so that the scenarios based
on heavy $\nu_\tau$ decaying into $\nu_\mu$ and some scalar particle
\cite{kawa}
are immediately ruled out.  Suppose then that the atmospheric anomaly is
due to
mixing between $\nu_\mu$ and some sterile neutrino.
Now $\nu_\tau$ can be heavy and having it decay away to muon neutrino and a
scalar state prior nucleosynthesis could alleviate the tension somewhat.
The effect is roughly equivalent to having about a half a neutrino degree of
freedom worth less energy density in the universe \cite{kawa} (less energy
density
leads to slower expansion, and hence later decoupling of $n/p$-ratio).
However, the
sterile state with the requested mixing parameters is brought into full
thermal
equlibrium due to oscillation and quantum damping prior to nucleosynthesis
\cite{ekt}, overcoming the alleviating effect discussed above and making the
tension even worse.  The only possibility to alleviate the tension is that
$m_{\nu_\tau} \lsim 1$ MeV and $\nu_\tau$
decays into an {\it electron neutrino} in the
short interval after the electron neutrino freezeout but prior the onset of
nucleosynthesis. In this case the excess (almost thermal) electron
neutrinos can
significantly increase the weak interaction rates keeping the $n/p$-ratio in
equilibrium longer and hence leading to much less helium being produced
\cite{eeff,dgt}.
Bringing the sterile neutrino into equilibrium makes also this solution less
effective, but is not strong enough to rule out the possibility entirely
\cite{dgt,steen}.

The chemical  evolution of D and $\UHe$ is particularily simple: $\UHe$
increases
with time, whereas D decreases.  In contradistinction, $\EHe$ and $\ZLi$
are both
produced and destroyed during galactic chemical evolution.  Thus it is much
more difficult to make reliable claims of their primordial abundances based
on present abundances.  Observed $\EHe$ abundances \cite{He3obs} in particular
vary a lot within the galaxy and $\EHe$ observations are useful for
constraining BBN only when combined with D and chemical evolution models.

For $\ZLi$ there is a very impressive plateau \cite{plateau} of abundances in
PopII stars with surface temperatures $T$ = 5800--6400 K. The observed
value is \cite{Molaro,PSW}
\beq
   \log_{10}(\ZLi/{\rm H}) = -9.75\pm0.10.
\label{molaro}
\eeq
The universality of this abundance suggests that it is closely related to the
primordial abundance.  There may have been some depletion, i.e., some of the
surface $\ZLi$ has been destroyed by the star.  Therefore the primordial
abundance should be larger by some depletion factor $D_7$.  Pinsonneault \etal
\cite{PSW} estimate $D_7$ = 0.2--0.4 dex. This corresponds to a primordial
\beq
   \log_{10}(\ZLi/{\rm H})_p = -9.45\pm0.20.
\label{pinso}
\eeq
However, Vauclair and Charbonnel \cite{VC98} give a lower estimate
\beq
   \log_{10}(\ZLi/{\rm H})_p = -9.65\pm0.10.
\label{vc}
\eeq
These estimates for lithium are compatible with either a low,
$\et \sim 1.5$, or a high, $\et \sim \hbox{4--6}$ baryon density, but disfavor
a compromise value $\et \sim \hbox{2.5--3}$.

A promising new method with the potential to resolve this $\eta$ dicothomy is
the observation  of deuterium in clouds at high redshifts by its absorption
of quasar light \cite{highz}. Some of these clouds are so far away, that when
the observed light passed through them, the universe was a mere one-tenth of
its present age; thus the matter in these clouds and therefore the observed
deuterium abundance should be close to primordial. Unfortunately, at present
we only have a small number of such D/H measurements, and even the existing
ones are still controversial. Burles and Tytler \cite{BT98} obtain from their
two best observations
\beq
   \DH = 3.4\pm0.3\times10^{-5},
\eeq
which corresponds to $\et = 5.1\pm0.3$ in SBBN. This is in contradiction with
the OSS97 estimate $Y_p \sim 0.23$.  However, the analysis of Burles and
Tytler has been debated \cite{lowDdebate} and one observation by HST
\cite{webbetal} from an absorption cloud at $z=0.7$ appears to give
a {\em high} value of $\DH \sim 2\times10^{-4}$.

Thus the observational situation remains unclear. If we suppose that
some of the determinations of primordial abundances are correct, but we do
not  know which, we are led to an SBBN range
\beq
   \et \sim \hbox{1.5--6}.
\eeq

One can also try to determine the universal baryon density by other means,
discarding nucleosynthesis considerations\cite{SHF}.  Determinations of this
kind are rather uncertain at present, but tend to favor the larger values
of $\eta$.

In conclusion, there is an unsettled disagreement between different
observations in the context of SBBN. While the problem may lie with the
observations, or in the determination of primordial abundances from them,
another possibility is, that the primordial abundances indeed do not correspond
to the same $\eta$ in SBBN, so that it needs to be modified. In this paper we
study the possibility of inhomogenous big bang nucleosynthesis (IBBN) in light
of the present observational situation.
In section II we
discuss the generic mechanisms known to produce inhomogeneities in the baryon
distribution and 
the significance of the distance scale of the inhomogeneity.
We describe our numerical calculations in section III and
give our results in section IV. Section V contains our conclusions.

\section{Inhomogeneous Nucleosynthesis}
\la{sec:ibbn}

In SBBN we assume baryonic matter was homogeneously distributed during
nucleosynthesis, but actually we do not know whether this was the case.
If the inhomogeneity was of sufficiently small scale, diffusion would
have homogenized the matter distribution before the formation of the
cosmic microwave background leaving no directly observable trace today.

\subsection{Generating the inhomogeneity}

Various phase transitions which took place before nucleosynthesis in
the early universe were capable of producing large-amplitude
small-scale fluctuations in the baryon number density: in particular the
electroweak (EW) transition at $T \sim 100\GeV$ and $t \sim 10^{-11}$ s
and the QCD transition at $T \sim 150 \MeV$ and $t \sim 10^{-5}$ s.

IBBN was studied extensively in the late 1980's, when it was realized that
a first-order QCD phase transition in the early universe could produce the
kind of inhomogeneity which would affect BBN
\cite{AHS87,AFM,MF88,KMCRW,TS89,KMOS,ADFMM,MMAF90,KM90,MAMF91}.
At first \cite{AHS87,MF88,TS89} it seemed possible
to accommodate much larger values of $\et$, even
$\Omega_b = 1$, but more detailed calculations
\cite{KMCRW,KMOS,MMAF90,TSOMMF,JFM94,MKO}
showed that the upper bound to $\et$ was in fact much less
increased.

The original mechanism relying on chemical pressure \cite{Witten},
operative in the QCD transition, leads to a geometry
where localized clumps of high density are surrounded by large voids of
low baryon density \cite{AH85,K88,IKKL94,CM}. The details of the QCD
transition are poorly known and both the amplitude and the size of the
inhomogeneities can vary significantly; the size of course is bounded by
the horizon at the QCD transition, which is about $2\times10^6
\hbox{ m (at } T = 1 \hbox{ MeV)} = 0.4 \hbox{ pc (today)}$.

Also the electroweak phase transition (EWPT)
generically produces inhomogeneities,
and possibly with large density contrasts. This assumes of course that
the baryons we see around us today, were generated during the electroweak
phase transition \cite{baryogen}. Some scenarios \cite{CPR,GS98}
may even give rise to regions of antibaryons mixed with the overall
baryonic excess, leading to the interesting possibility
of nucleosynthesis in the presence of antibaryons \cite{Rehm98,HannuElina}.
The generic feature leading to the formation of inhomogeneities in the more
standard scenarios is the strong dependence of the baryoproduction rate
on the bubble wall velocity in the so called "charge transport mechanism"
\cite{CKN,FS,CKV}, coupled with the characteristic changes in the velocity
of the bubble walls during the transition \cite{Heckler}. For thin walls one
finds a local baryoproduction rate
\beq
  B(x) \approx c/v_{\rm w}(x).
\label{bmax}
\eeq
The velocity dependence of the local baryoproduction rate due to the
"classical chiral force" mechanism \cite{JPT,CJK1}, operative in the limit
of wide walls, is much weaker \cite{CJK2}.  However, the generic geometry
of inhomogeneities arising from EWPT is quite opposite to the QCD case;
voids of low density surrounded by walls of high density.

After nucleation bubble walls quickly accelerate to a terminal velocity
$v_{\rm w} \sim 0.1\mbox{--}0.5c$, whose exact value depends on the parameters
of the phase transition, like the latent heat released, the surface tension
and the frictional forces effected on the bubble wall by the ambient plasma
\cite{MP,KL96}.  After some time (we are only considering deflagration
bubbles here), the shock waves preceding phase transition fronts collide
reheating the unbroken phase plasma back to the critical temperature.
As a result the pressure forces driving the bubble expansion are reduced,
and, were it not for the general expansion of the universe, the walls
would become to a complete stop.  Due to Hubble expansion the walls can
still continue expanding, but now with a greatly reduced speed, typically
$v_{\rm w} \sim {\cal O} (\mbox{few}) \times 10^{-3}c$ \cite{KL96}.
These velocity scales and the rate (\ref{bmax}) indicate that the maximal
density contrast possibly generated by the EW mechanism is about $\sim 100$.

The typical size of the voids in this ``beer foam" geometry is some fraction
of the horizon at the EW transition, $\ell_H = 3\times10^3 \hbox{ m (at 1
MeV)} = 6\times10^{-4} \hbox{ pc (today)}$.  A nucleation calculation, which
ignores the thermodynamics of the bubble interactions, typically gives for
the size of bubbles at the coalescence only $\ell_{b} \sim 10^{-3} \ell_H$
\cite{KL96,FJMO94,LR98}.
However, due to reheating the firstly nucleated bubbles may
inhibit the growth of bubbles formed only slightly later, increasing perhaps
significantly the size of the largest structures as compared with the
simplest nucleation estimate. Also in extended scenarios including magnetic
fields \cite{GS98}, the size of a single bubble can reach the horizon scale.
We then consider the inhomogeneity size a free parameter, with values
$r \sim 10^{-3}-1 \ell_H$.

\subsection{Distance scales}

Both the EW and the QCD transition appear capable of producing
high initial
density contrasts. In both cases the density fluctuations would
be non-gaussian, consisting of high- and low-density regions.
The pattern would not be regular, but it would have a characteristic
distance scale.
The inhomogeneity can be described by the typical
geometric shape of these regions and the following three parameters:
 1) typical distance scale $r$,
 2) typical density contrast $R \equiv \eta_{\rm high}/\eta_{\rm low}$,
and 3) the volume fraction $f_v$ of the high-density regions.

The distance scale $r$ is especially important. An inhomogeneity can have a
large  effect on nucleosynthesis only if the distance scale is comparable to
the neutron diffusion length $d_n$ during nucleosynthesis.
If the distance scale is too small, $r \ll d_n(500 \keV) \sim 200 \hbox{ m
(at 1 MeV)} \sim 4\times10^{-5} \hbox{ pc (today)}$, the inhomogeneity is
erased before nucleosynthesis, because before the weak freeze-out protons and
neutrons are constantly converted to each other by weak reactions, and the
diffusion thus evens out both the proton and the neutron density.

If the distance scale is large, $r \gg d_n(10 \keV) \sim 500 \hbox{ km
(at 1 MeV)} \sim 0.1 \hbox{ pc (today)}$, diffusion does not occur until
nucleosynthesis is completed.  In this ``ordinary inhomogeneity'' scenario,
the high- and low-density regions undergo independent standard BBN with
$\eta_{\rm high}$ and $\eta_{\rm low}$ and the matter is mixed afterwards
to have the average baryon-to-photon ratio $\eta$.  Leonard and Scherrer
\cite{LS96} have shown that this kind of inhomogeneity cannot increase
the upper limit to $\eta$, since the inhomogeneity raises $\UHe$\ yields.
Arbitrarily low $\eta$ can be made acceptable with these models, however.

An intermediate distance scale, $d_n(500\keV) < r < d_n(10\keV)$ leads to
nucleosynthesis with inhomogeneous neutron-to-proton ratio. The strongest
effect occurs if $n/p \gg 1$ in some region, because in this region neutrons
are left over from $\UHe$ synthesis. This can be induced in the low-density
region if $f_vR \gg (n/p)^{-1}$ before diffusion.
QCD-scale inhomogeneities could be\cite{IKKL94} of the scale required,
although QCD lattice calculations \cite{QCDlattice} favor values below the
short end of this range. For the EW case this range corresponds to
a fluctuation scale $r \gsim 0.1 \ell_H$ during the transition.

There may be other possible sources of baryon inhomogeneity in addition
to the EW and QCD phase transitions. Moreover, there is a considerable
uncertainty regarding the parameters $r$, $R$ and $f_v$ from each
transition. Therefore it is natural to treat the two questions separately:
1) Are there IBBN parameter regions where IBBN agrees with observations
equally well or better than SBBN?  2) Could the EW or QCD phase transitions
produce inhomogeneity in this parameter region?

\section{Computations}
\la{sec:com}

The IBBN code used for this paper is based on the code used in \cite{KM90}
and the nuclear reaction rates have been updated according to \cite{SKM93}.
In the $\UHe$\ yield we take into account the various corrections to the weak
reaction rates\cite{Dicus,KK94,Lopez}.
Theoretical uncertainty in abundance yields due to uncertainty in nuclear
reaction rates is usually small compared to observational uncertainties.
An exception is the $\ZLi$ yield for which one standard deviation is
0.07--0.10 dex upwards and 0.11--0.19 downwards \cite{SKM93}. We take this
into account when obtaining limits on $\eta$ from $\ZLi$ deduced from
observations, by further relaxing the upper
limit to $\ZLi$ by 0.15 dex and the lower limit by 0.10 dex. Proton diffusion
is included according to \cite{Jedamzik}. The convergence of the code has been
improved by combining the nuclear reaction and diffusion steps into a single
step, which reduces the number of time steps needed for accurate results.

We assume spherical symmetry and use a nonuniform
radial grid of 64 zones representing
a sphere with comoving radius $r$, with reflective boundary conditions both
at the center and at $r$. This setup allows us to model both geometries
discussed above: assuming centrally condensed density describes the QCD-type
and spherical shells of high density describe the EW-type geometry. The volume
fraction covered by the high density region in each geometry is
\bea
   f_v & = & f_r^3   \mbox{\qquad\qquad\qquad (centrally condensed)}, \\
   f_v & = & 1-(1-f_r)^3    \mbox{\qquad (spherical shell)},
\eea
where $f_r$ denotes the fraction of the radius covered by
the high-density region. Given the geometry, the model is
specified by four parameters: $r$, $f_v$, $R$ described above, and the average
baryon-to-photon ratio $\eta$. The numerical value for $\eta$ always refers
to the present value, i.e., after ${\rm e}^+{\rm e}^-$-annihilation.

Note that these inhomogeneities are
in baryon number only.  At nucleosynthesis time, the energy density is
dominated by radiation by a factor of at least $10^5$.  The density of
baryon number can hence be strongly inhomogeneous without a noticeable
dynamic effect, and the main process through which the inhomogeneity
evolves after the phase transition is diffusion \cite{AHS87,Jedamzik,JFM94}.

The effect of neutron diffusion is to reduce $\UHe$\ and $\ZLi$\ yields, and
to increase D and $\EHe$\ yields. All these changes are in the direction of
favoring a larger $\eta$.  However, diffusion has to compete with the ordinary
inhomogeneity effect which for $\UHe$, $\ZLi$, and $\EHe$\ is the opposite,
increasing $\UHe$\ and $\ZLi$\ and reducing $\EHe$.  For D this latter
effect depends
on the average $\eta$.  For small $\eta$, D is reduced, and for large $\eta$ D
is increased.

The most dramatic effect is obtained when the neutron diffusion out of the
high-density region leads to a large excess of neutrons in the low-density
region. This requires a density contrast
\beq
   R \gg \biggl(\frac{p}{n}\biggr)_0\frac{1}{f_v},
\eeq
where $(p/n)_0 \sim 7$ is the SBBN proton/neutron ratio at the onset of
nucleosynthesis. Increasing $R$ leads to a stronger effect, but the increase
soon saturates. Indeed, for large $R$ almost all of the nuclear matter already
is in the high-density region, while almost no matter remains in the low
density regions:
\bea
  \eth & = & {{R\eta}\over{f_vR+1-f_v}} 
    \stackrel{R\rightarrow \infty}{\rightarrow}
    \frac{\eta}{f_v}, \\
  \etl & = & {{\eta}\over{f_vR+1-f_v}} \sim \frac{\eta}{f_vR}
    \stackrel{R\rightarrow \infty}{\rightarrow} 0.
\eea
\begin{table}[tbh]
\begin{tabular}{rllr}
R &  $f_r$ & $f_v$ & $f_vR$ \\ \hline
\multicolumn{4}{c}{Centrally condensed (c.c.)} \\ \hline
  283 & $1/\sqrt{2}$ & 0.3536 & 100 \\
  800 & $1/2$        & 0.125  & 100 \\
 2263 & $1/2\sqrt{2}$& 0.0442 & 100 \\
 6400 & $1/4$        & 0.0156 & 100 \\
51200 & $1/8$        & 0.0020 & 100 \\ \hline
\multicolumn{4}{c}{Spherical shell (s.s.)} \\ \hline
 1000 & 1/4   & 0.5781 & 578 \\
 1000 & 1/8   & 0.3301 & 330 \\
 1000 & 1/16  & 0.1760 & 176 \\
 1000 & 1/32  & 0.0909 &  91 \\
 1000 & 1/64  & 0.0461 &  46 \\
 1000 & 1/128 & 0.0233 &  23 \\
\end{tabular}
\vspace*{3mm}
\caption[a]{\protect
The different geometries studied.
$R$ is the
density contrast between the high and low density, $f_r$ is the high-density
fraction of the grid radius, and $f_v$ is the corresponding volume fraction.}
\end{table}
The effect of further increasing $R$, beyond, say $R = 100/f_v$, just leads
to a further reduction of matter density in the low-density region and has
essentially no effect on nuclear yields.  An exception to
this may be D, since the D yield drops so fast with increasing $\eta$, that a
significant part of the D yield may still come from the low-density region,
giving rise to sensitivity to a reduction of $\etl$ and hence to $R$.

In most cases we chose to run with large enough $R$ to have close to this
maximal effect.  This leaves us with three parameters $f_v$, $\eta$, $r$.
We did runs with 11 different values of $f_v$ altogether (Table I).

For the runs with spherical shell geometry, we kept $R = 1000$ constant.
For the centrally condensed geometry some of the volume fractions were so
small, that a larger $R$ was needed to get the large inhomogeneity effect.
For the centrally condensed runs we kept the product
$f_vR = 100$ constant instead.

\section{Results}

It has been customary in IBBN studies \cite{KMOS,MAMF91,TSOMMF,MKO} to
plot the regions in the $(\eta, r)$-plane allowed by different observational
constraints. Since the observational situation has become rather less clear
recently, we present the results first
as abundance contours for a given geometry and
$f_v$, so different constraints can then be applied afterwards.
For $\UHe$\ we plot the mass fraction $Y_p$, for D and $\ZLi$\ we plot the
number ratios D/H and $\ZLi$/H. To save space, the less interesting $\EHe$
is not shown.

\begin{figure}[h]
\vspace*{-0.4cm}
\epsfysize=6.8cm
\epsffile{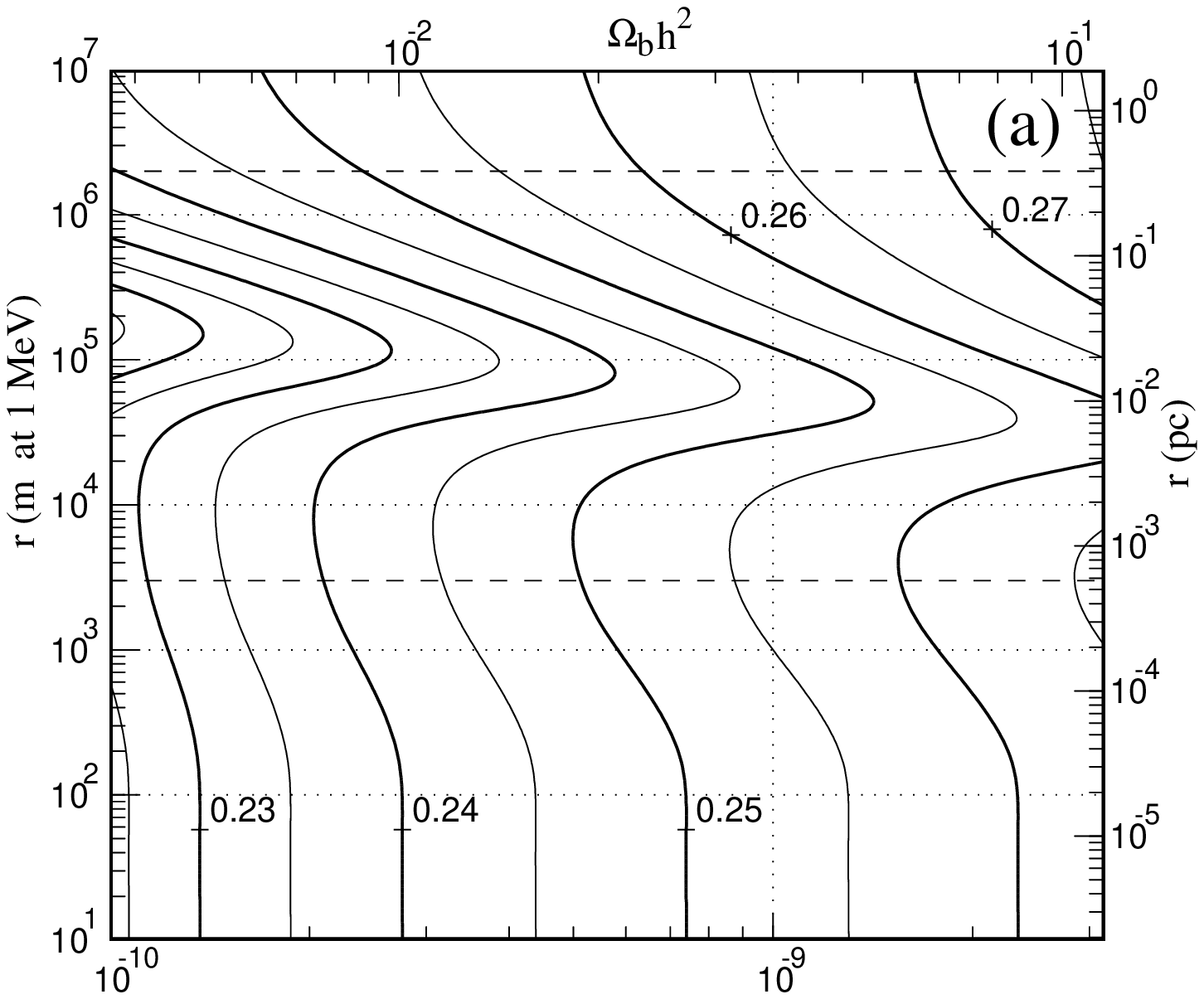}
\vspace*{-0.1cm}
\epsfysize=6.8cm
\epsffile{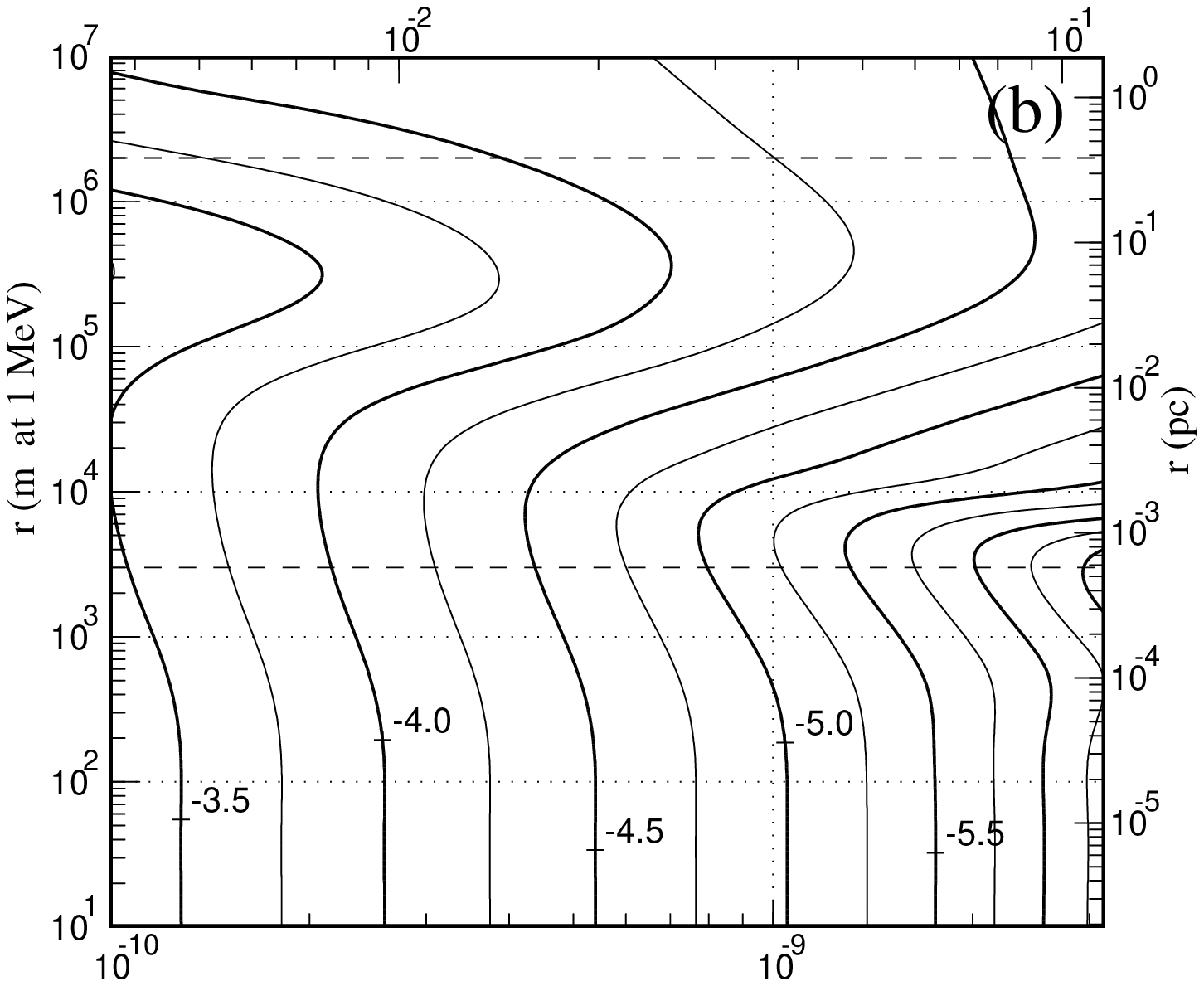}
\vspace*{-0.1cm}
\epsfysize=6.8cm
\epsffile{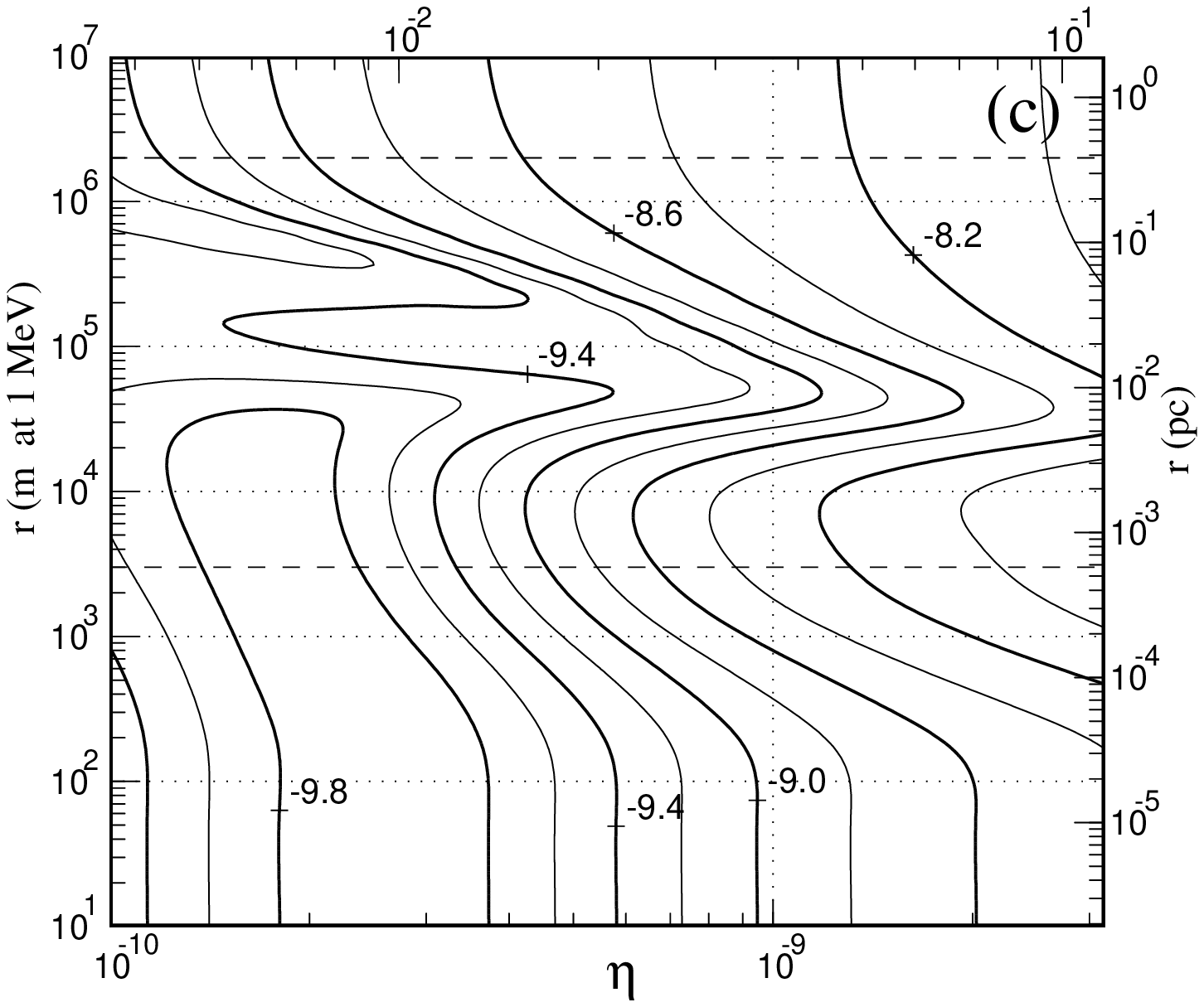}
\caption[a]{\protect
The $\UHe$, D, and $\ZLi$\ yields from inhomogeneous nucleosynthesis runs with
the spherical shell geometry, with $R = 1000$ and $f_r = 1/16$ ($f_v = 0.176$).
The contours of (a) $Y_p$, (b) $\log_{10} \DH$,
and (c) $\log_{10} \ZLiH$ are plotted as
a function of the average baryon-to-photon ratio $\eta$ and the distance scale
$r$ of the inhomogeneity.  The two horizontal dashed lines denote the horizon
scale $\ell_H$ at the QCD (upper) and EW (lower) phase transitions.
}
\label{S16}
\bigskip
\end{figure}

\begin{figure}[h]
\epsfysize=15.0cm
\epsffile{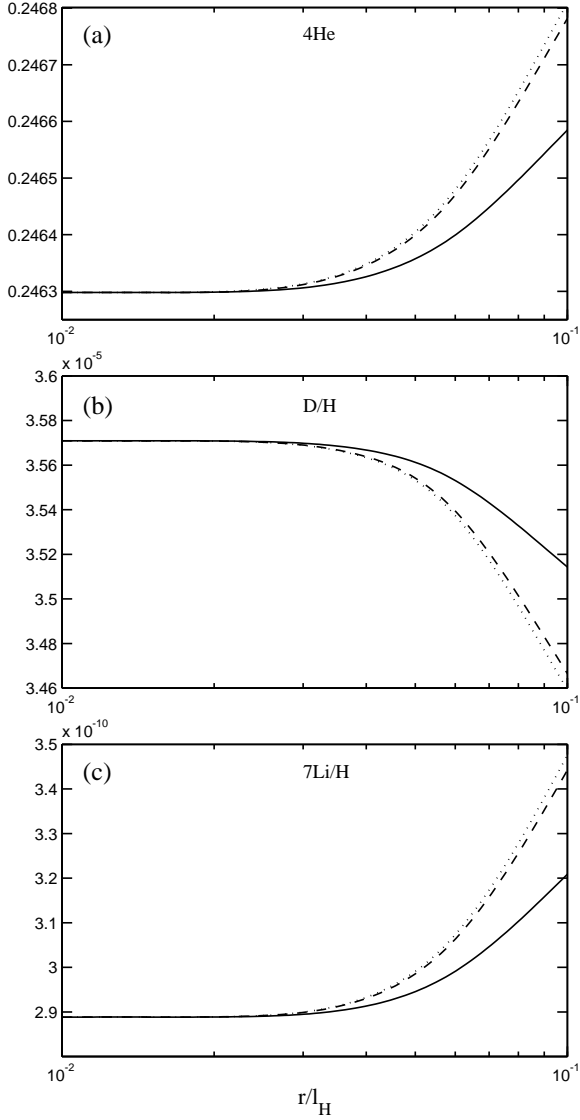}
\caption[a]{\protect
Effects of small scale inhomogeneity on the isotope yields.  This figure is for
the spherical shell geometry appropriate for the EW transition, and for $f_v =
0.3301$ and $\et = 5$.  The three lines correspond to $R = 10$ (solid),
100 (dashed), and 1000 (dotted)
(or $f_vR = 3.3, 33, 333$), showing how the effect saturates for large $R$,
so that there is little difference between $R = 100$ and $R = 1000$.
The horizontal axis gives the ratio of the distance scale to the EWPT
horizon.
}
\label{smallfig}
\bigskip
\end{figure}

In Fig.~\ref{S16} we show the results for the spherical shell (s.s.) geometry
with $R = 1000$ and $f_r = 1/16$.  It is clear that at the distance scales
attainable in the EW transition
(indicated by the lower horizontal dashed line in
the figure) the IBBN results do not significantly differ from SBBN results;
the observational uncertainties are certainly much larger.  However, even
with scales as small as $r \sim 0.05\ell_H$, the effect of inhomogeneity
(see Fig.~\ref{smallfig}) can be larger than certain small corrections
recently included into the SBBN computations \cite{Lopez}.

\begin{figure}[tbh]
\vspace*{-0.4cm}
\epsfysize=6.8cm
\epsffile{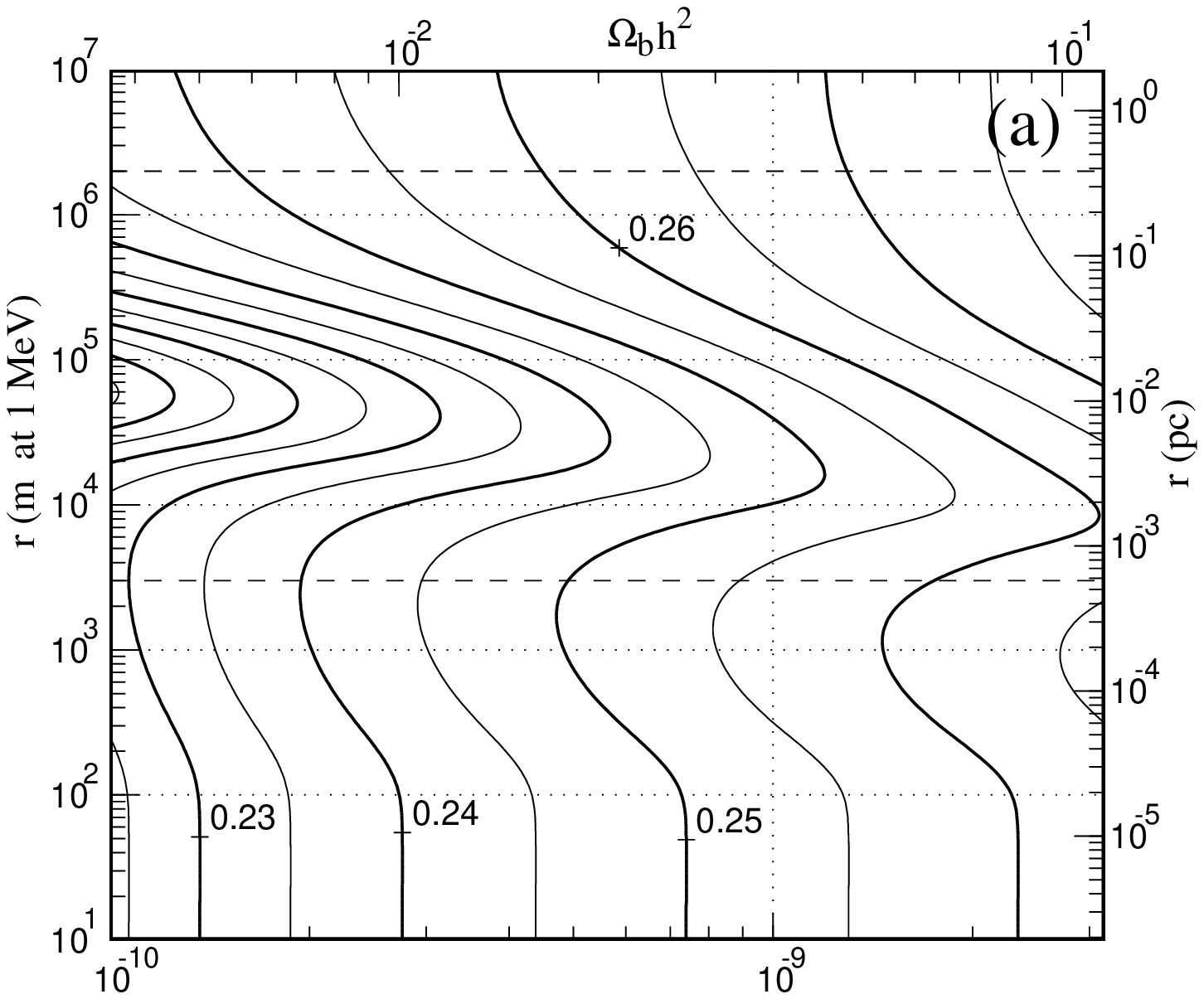}
\vspace*{-0.1cm}
\epsfysize=6.8cm
\epsffile{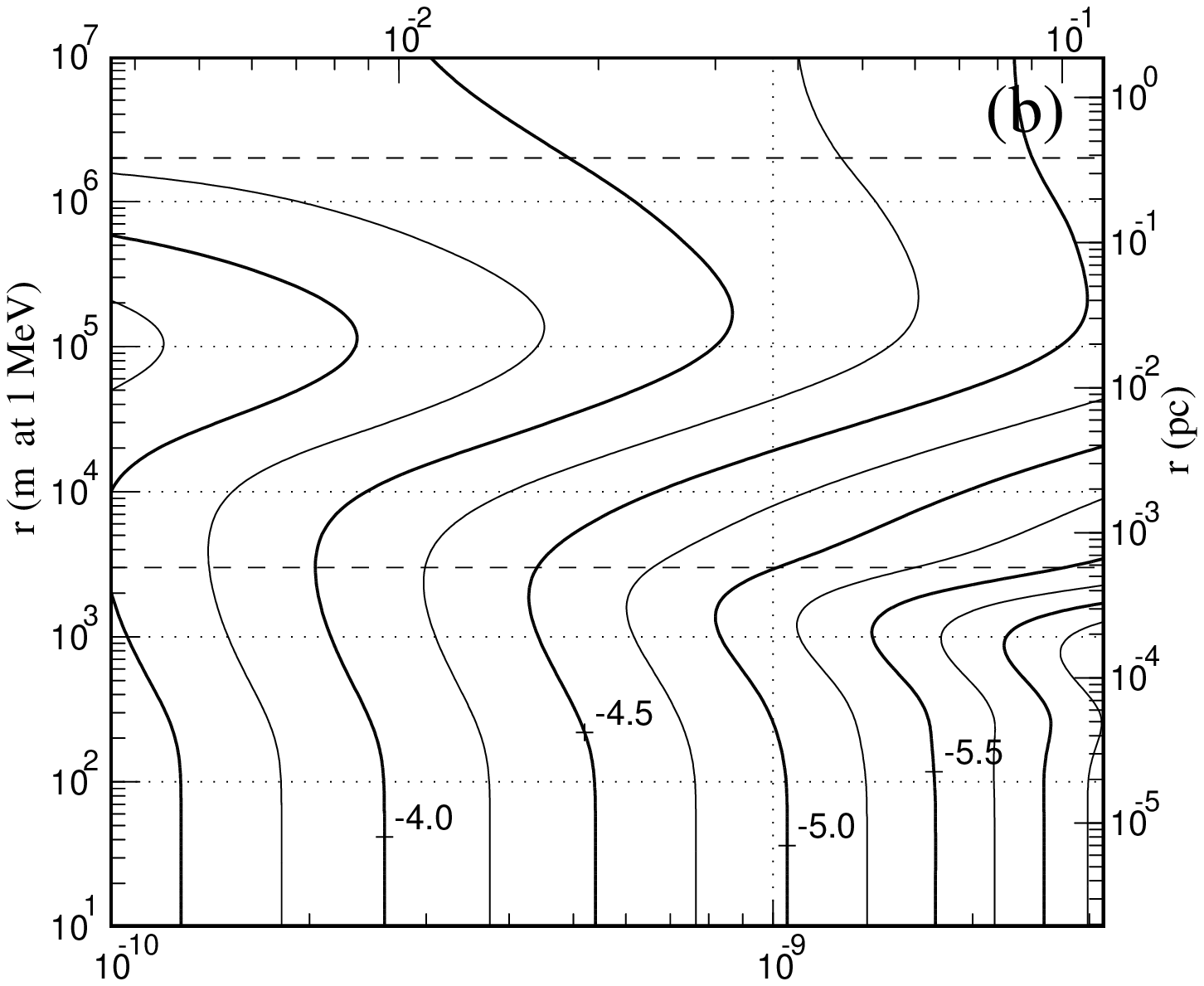}
\vspace*{-0.1cm}
\epsfysize=6.8cm
\epsffile{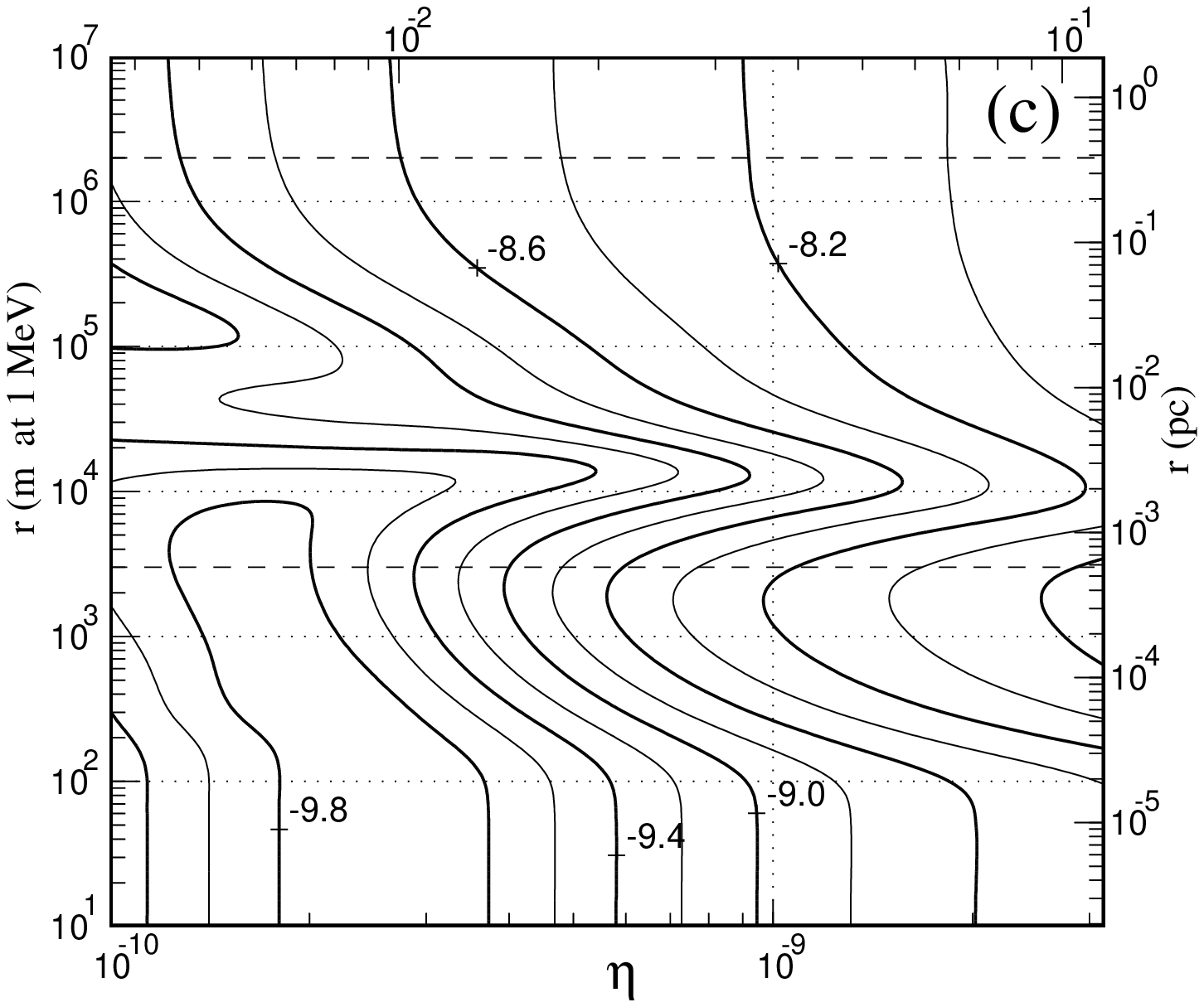}
\caption[a]{\protect
  Same as Fig.~\ref{S16}, but for the
centrally condensed geometry with $R = 800$
and $f_r = 1/2$ ($f_v = 0.125$).
}
\label{C2}
\end{figure}

\begin{figure}[tbh]
\vspace*{-0.4cm}
\epsfysize=6.8cm
\epsffile{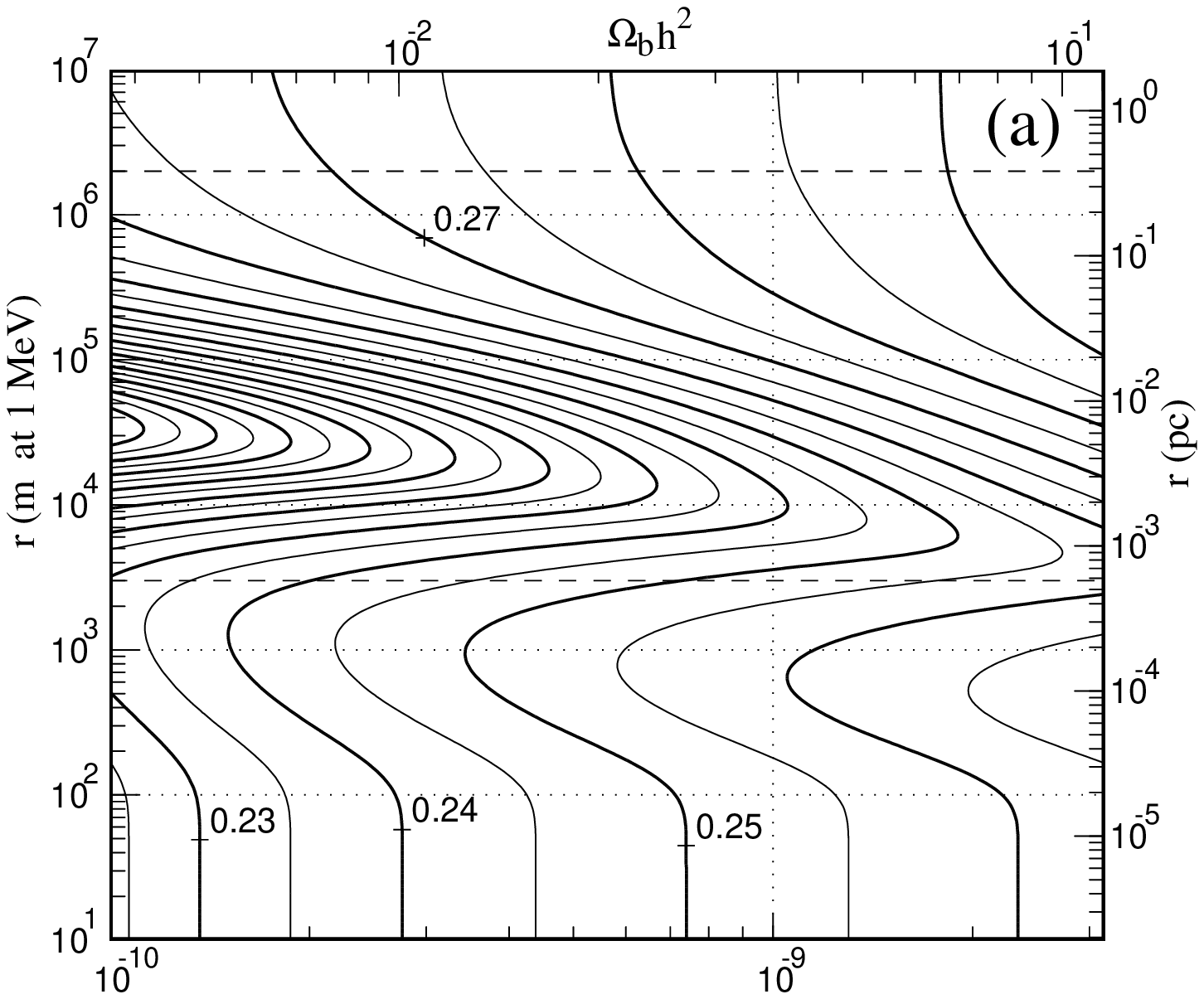}
\vspace*{-0.1cm}
\epsfysize=6.8cm
\epsffile{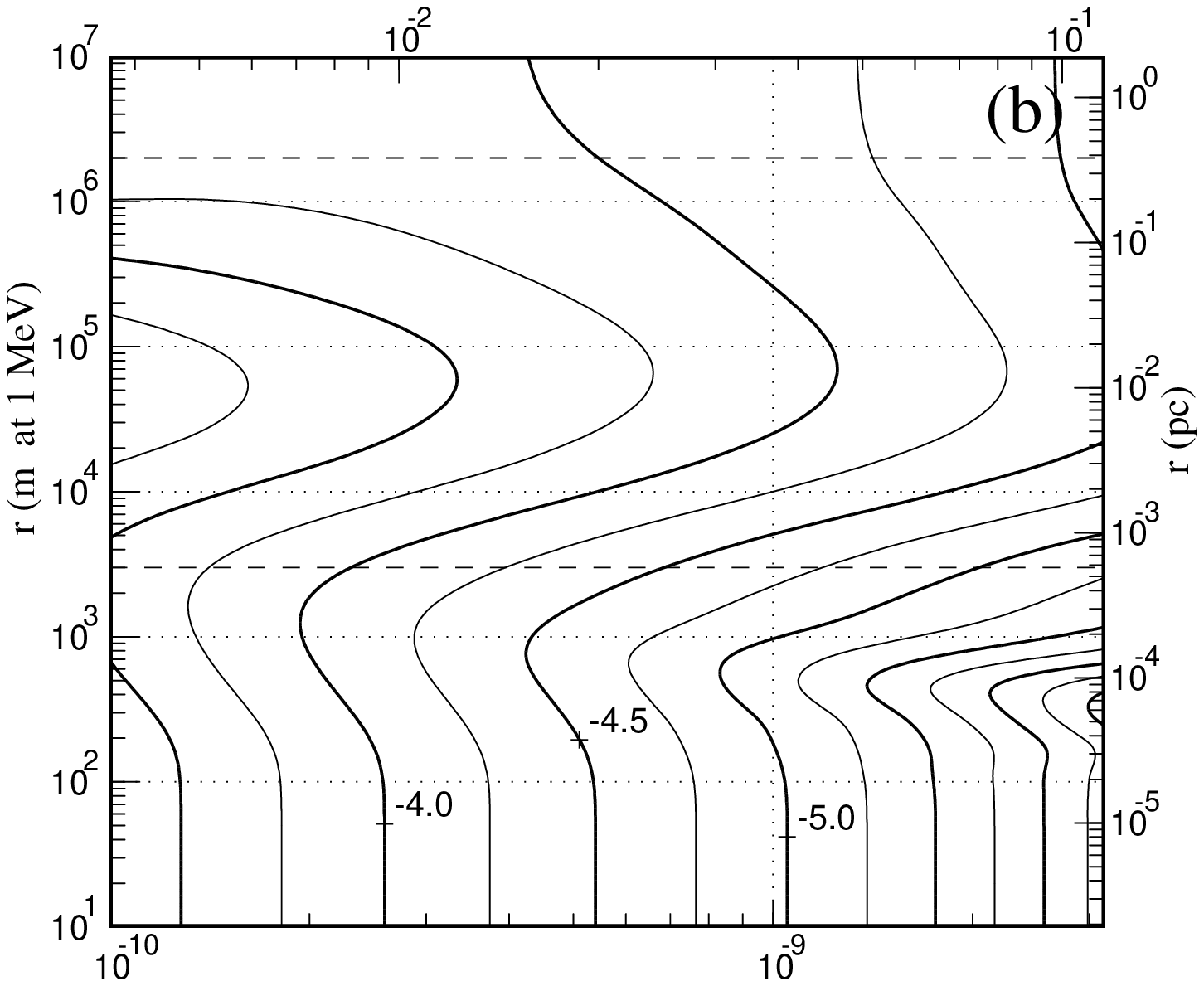}
\vspace*{-0.1cm}
\epsfysize=6.8cm
\epsffile{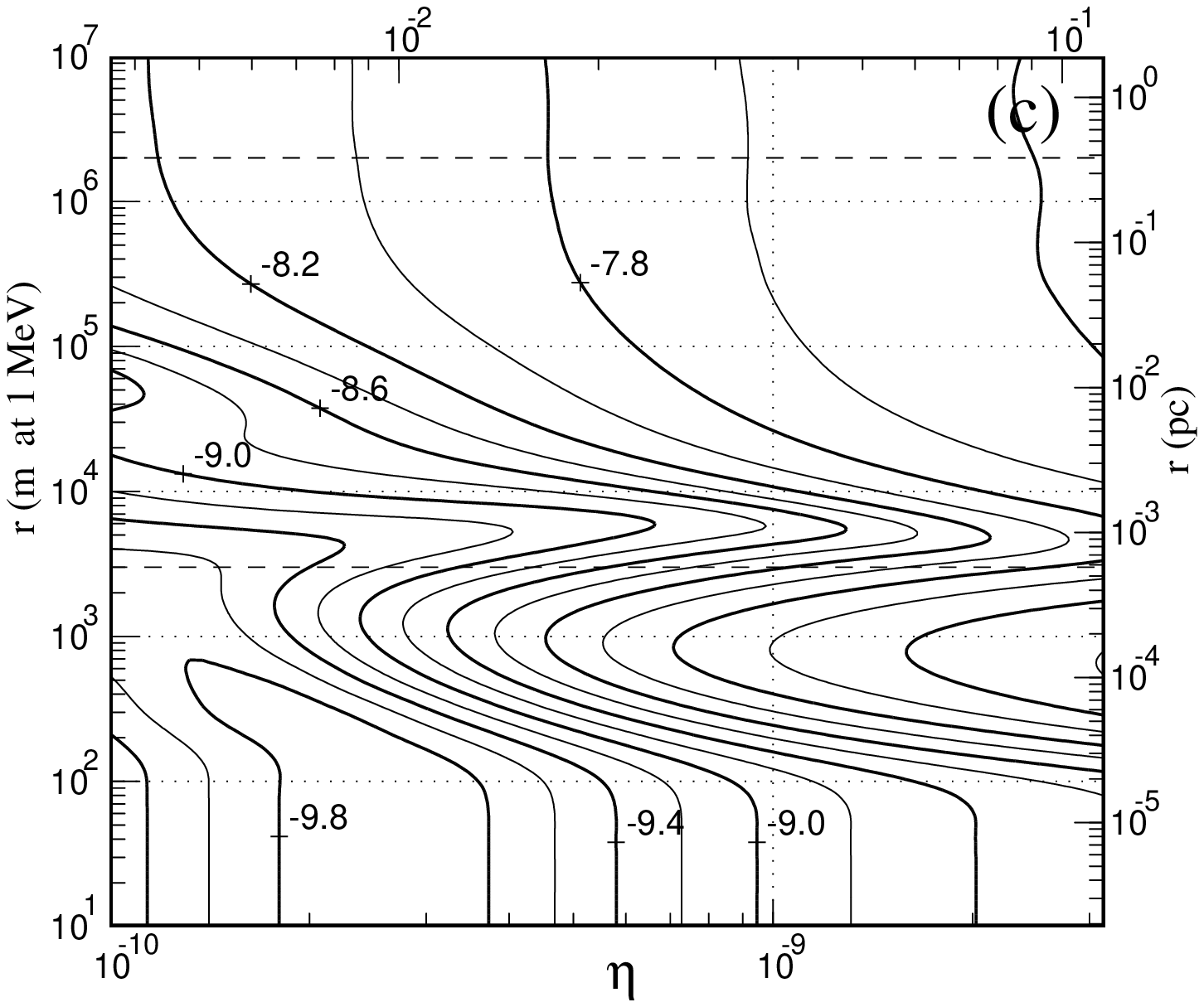}
\caption[a]{\protect Same as Fig.~\ref{S16}, but for the
centrally condensed geometry with $R = 6400$ and $f_r = 1/4$
($f_v = 0.0156$).}
\label{C4}
\end{figure}

Now we take a somewhat different point of view and consider
a broader range of density contrasts and
distance scales than can be produced in the electroweak phase
transition. In Figs.~\ref{C2} and \ref{C4} we present the results for the
centrally
condensed (c.c.) runs with $f_r = 1/2$ and $f_r = 1/4$. The results from other
runs described in Table I are qualitatively similar.

\subsection{Optimum scales}

How much $\UHe$ is produced depends on the number of neutrons available.
The yield is minimized at an optimum distance scale $\ropt \sim 10^4 - 10^5$
m, where a maximal number of neutrons diffuse out from the high-density
region (where most of the $\UHe$ is produced), but not too many of them
diffuse back when the nucleosynthesis in the high-density region starts
consuming free neutrons, and the direction of the neutron diffusion
reverses.  D yields are maximized at scales somewhat larger than $ \ropt$,
in particular for large $\eta$, because of the strong ordinary inhomogeneity
effect.  Situation is more complicated with the $\ZLi$\ yields, but they
tend to be minimized at $r \lsim \ropt$.

We find that $\ropt$ goes down with increasing $\eta$, roughly as
$\eta^{-2/3}$.  Also, the s.s.\ geometry gives a larger optimum scale than
does the c.c.\ geometry with the same $f_v$, and the dependence on $f_v$ is
different with different geometries: for centrally condensed spheres $\ropt$
goes down with decreasing $f_v$, whereas for spherical shells it increases
with  decreasing $f_v$.

It is possible to derive the parametric dependence of $\ropt$ on $\eta$
and  $f_v$ analytically. Consider the diffusion of neutrons after the weak
freeze-out but before the start of nucleosynthesis. The flux of neutrons
into the low-density region is proportional to the neutron diffusion
coefficient $D$, to the surface area $A$ of the boundary, and to the
gradient of the neutron density at the boundary, roughly
$(\nhigh-\nlow)/\diflen$.  Here $\sqrt{Dt}$ is the diffusion length of
neutrons, and $\nhigh$ and $\nlow$ are the average neutron densities in
the high- and low-density regions, respectively. $\nhigh$ decreases as
\beq
   V{\partial n_{\rm high}\over \partial t} \sim
   -AD{\nhigh-\nlow\over\sqrt{Dt}}
\la{diffeq}
\eeq
where $V$ is the volume of the high-density region. If we ignore nuclear
reactions and weak inteactions, we can integrate out Eq.~(\ref{diffeq}).
Remembering that $f_v\nhigh+(1-f_v)\nlow=\nmean$ one readily finds that
the density contrast vanishes exponentially due to diffusion:
\beq
\nhigh-\nlow \sim\exp\left( -{A\over V}{\diflen\over (1-f_v)} \right).
\eeq
The optimum scale corresponds to
\beq
{A\over V}{\diflenns\over (1-f_v)} \sim 1
\la{condition}
\eeq
where $t_{\rm ns}$ is the starting time of nucleosynthesis. At scales
larger than the optimum scale, the neutrons have not diffused out
effectively before the synthesis of $\UHe$ begins. On the other hand,
making the scale smaller than the optimum scale does not significantly
increase the number of neutrons diffusing out, but makes the
back-diffusion at later times more effective.

Now it is easy to see why the optimum scales are smaller for condensed
spheres. For the same $f_v$, the surface-to-volume ratio $A/V$ is
smaller for condensed spheres than for shell geometry, which makes the
out-diffusion less effective and optimum scales smaller.

The efficiency of the out-diffusion also depends significantly on the
volume of the low-density region (term $(1-f_v)$ in the denominator of
Eq.~(\ref{condition})). If $f_v$ is large,
$\nlow$
increases rapidly, bringing the diffusion to end sooner than in the case of
small $f_v$.

The $\eta$-dependence of the optimum scale is through the
dependence on the diffusion length. The diffusion at the boundary
is controlled by the smaller diffusion coefficient of the high-density
region. The diffusion is dominated by scattering on protons,
$D_{np} < D_{ne}$. After electron-positron annihilation the diffusion constant
depends on the proton density and temperature as
\beq
   D_n \simeq D_{np} \propto {1\over\eth T^{5/2}}
  \approx {f_v\over\eta T^{5/2}}.
\eeq
The early universe expands as $t\propto 1/T^2$. The diffusion length at the
beginning of nucleosynthesis should go as
\beq
   \diflenns\propto\eth^{-1/2} T_{\rm ns}^{-9/4}
\eeq
The starting temperature $T_{\rm ns}$ of nucleosynthesis depends on $\eth$:
in higher density nucleosynthesis begins earlier. The dependence in the range
$\eth=10^{-10}\cdots 10^{-8}$ is $T_{\rm ns}\propto\eth^\gamma$,
$\gamma=0.07\cdots 0.1$. The diffusion length should depend on $\eth$ as
\beq
   \diflenns \propto \nhigh^{-1/2-9\gamma/4}
   =\nhigh^{-\alpha},
\eeq
where $\alpha = 0.65\mbox{--}0.73 \sim 2/3$, so that for the optimum scale
\beq
   (1-f_v){V\over A}\propto\eth^{-2/3}.
\la{hold}
\eeq
The surface-to-volume ratio of the high-density region is
\bea
\la{AV}
   {A\over V} & = &{3\over f_v^{1/3}r}
            \qquad \qquad \hbox{(c.c)} \\
   {A\over V} & = &{3\over r} {(1-f_v)^{2/3}\over f_v}
            \qquad \hbox{(s.s.)}. \nonumber
\eea
Combining equations (\ref{hold}) and (\ref{AV}) we find the observed
behaviour for the optimum scale
\bea
    \ropt
        & \propto & {f_v^{1/3}\over (1-f_v)} \eta^{-2/3}
        \qquad \hbox{(c.c.)} \\ 
    \ropt
        & \propto & {\eta^{-2/3} \over f_v^{1/3}(1-f_v)^{1/3}}
        \qquad \hbox{(s.s.)}. \nonumber
\eea

\subsection{Effects on the different isotopes}

$^4${\bf He.} For distance scales near the optimal one $\UHe$\ yields are
reduced.  The range in $r$ where the $\UHe$\ yield is below the SBBN value
covers 1--1.5 orders of magnitude.  For the optimum distance scale a given
$\UHe$\ yield is obtained for a value of $\eta$
which can be as much as four times
larger than in SBBN.   Making $f_v$ smaller causes a deeper reduction in
$\UHe$.  For the spherical shell geometry we get the maximum effect at
$f_r \sim 1/64$ although this may be due to our keeping $R$ fixed to $R =
1000$.  In the centrally condensed runs where we keep $f_vR$ constant
instead, the effect keeps getting stronger for smaller $f_v$.

{\bf D.}  The runs near the optimum scale produce more D than SBBN. At these
scales the D contours are pushed towards larger $\eta$ by about the same amount
as the $\UHe$\ contours. The maximum effect of the inhomogeneity is however at
larger scales. Both D and $\UHe$\ consistently allow larger $\eta$ in IBBN
near the optimum distance scales.

$^7${\bf Li.} Neutron diffusion helps to reduce the $\ZLi$\ yield, since at
late times neutrons are diffusing into the high-density regions and destroying
$^7{\rm Be}$\ there. However, the ordinary inhomogeneity effect is to increase
$\ZLi$\ yields, and usually this effect wins, but for some cases we get a net
reduction.

\subsection{Constraints on $\eta$}

We now compare our IBBN yields to observational constraints. Since at present
there is no agreement about what constraints to use, we consider a
number
of different sets of constraints.

 The most fundamental abundance
constraints are
the upper limit to primordial $\UHe$\ and the lower limit to primordial D/H,
obtained directly from observed abundances, since chemical evolution always
increases the $\UHe$\ abundance and reduces D/H. So, in our first set we
conservatively take for $\UHe$\ the 2-$\sigma$ upper limit
by Izotov and Thuan \cite{IT98},
\beq
   Y_p \leq 0.248
\la{Ypcons}
\eeq
and for D/H we use the present ISM abundance \cite{ISMobs} as the lower limit,
\beq
   \DH \geq 1.5\times10^{-5}.
\la{DHism}
\eeq

\begin{figure}[tbh]
\vspace*{-0.3cm}
\epsfysize=8.2cm
\epsffile{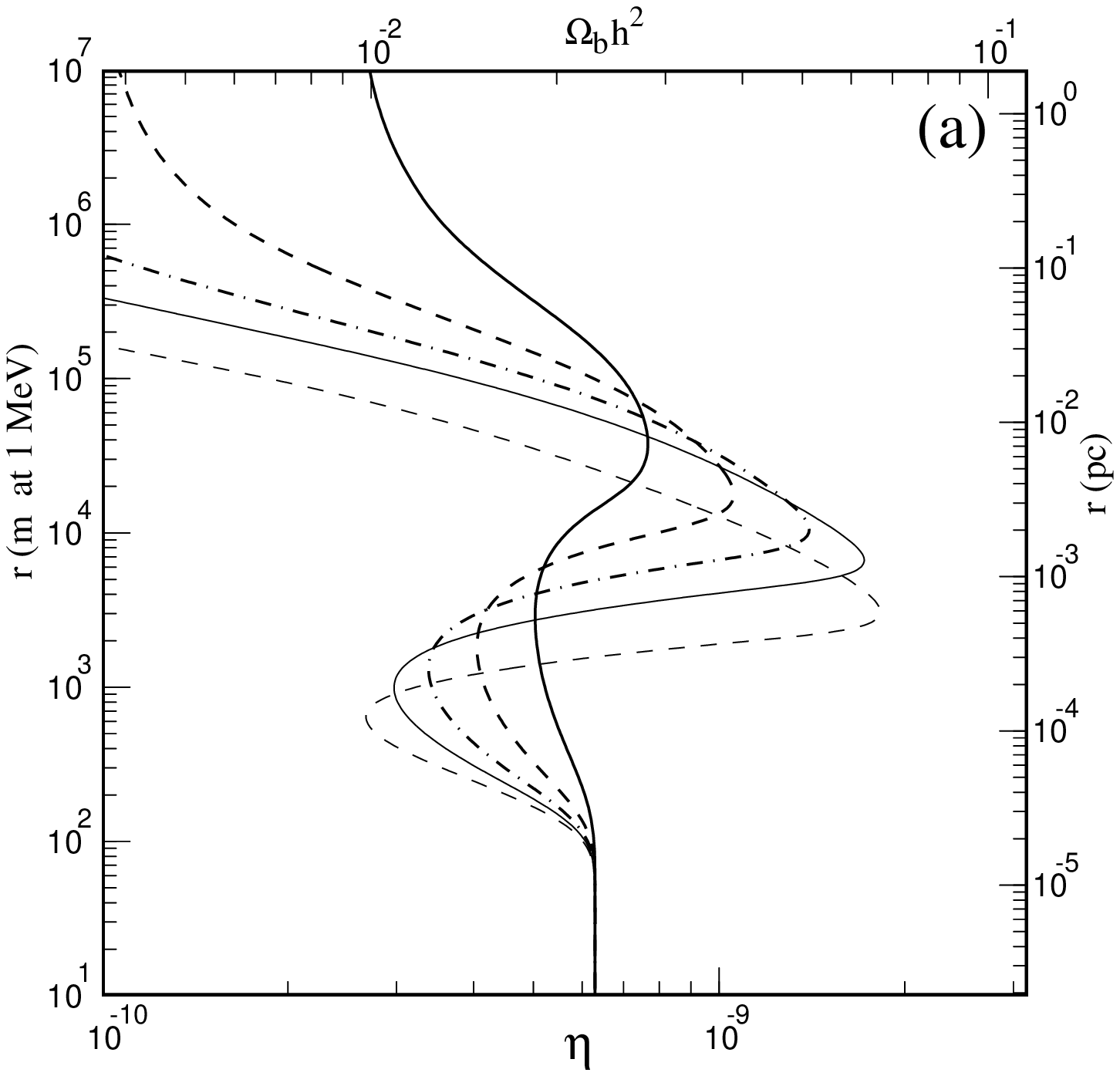}
\vspace*{-0.1cm}
\epsfysize=8.2cm
\epsffile{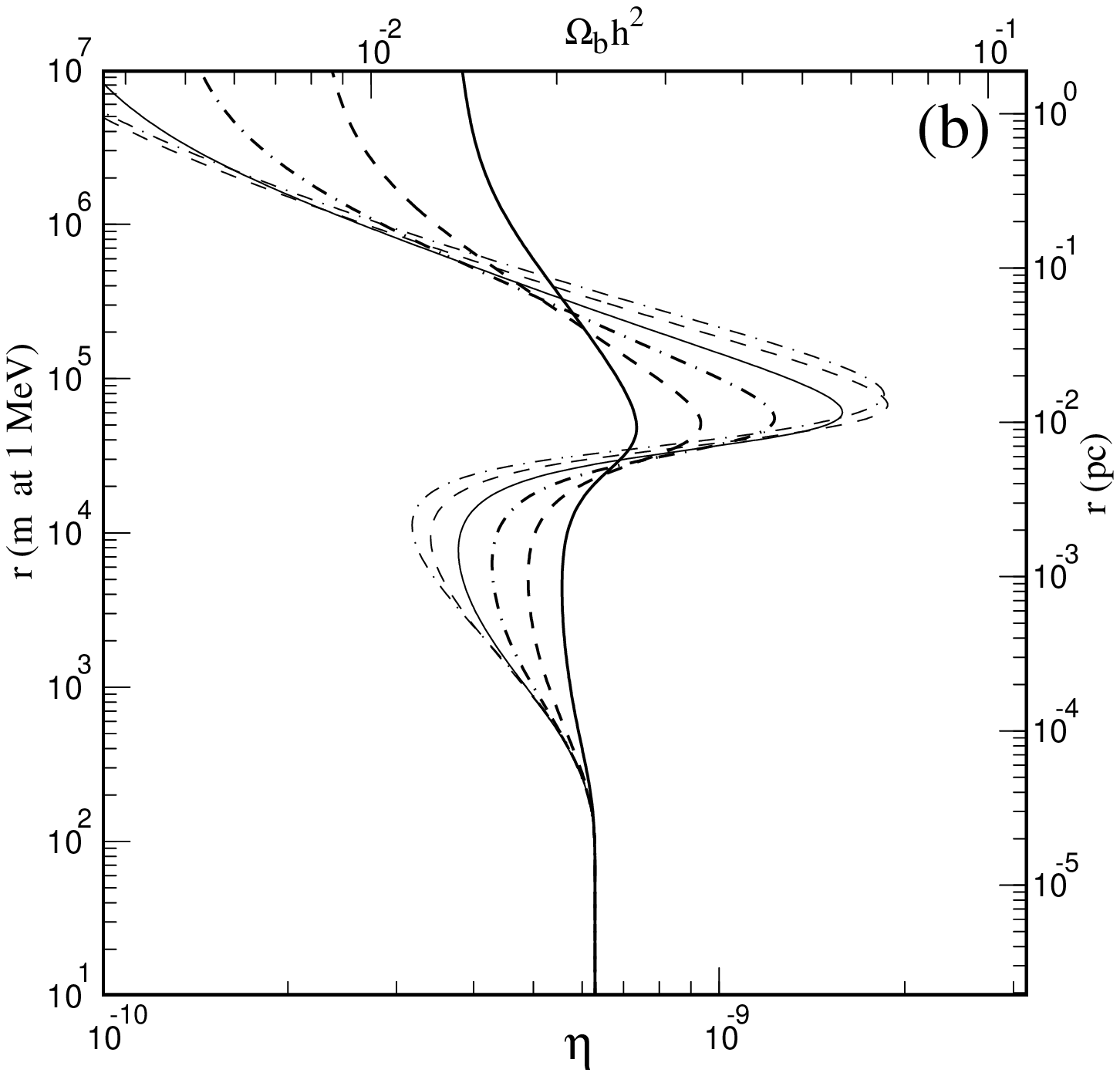}
\caption[a]{\protect
Conservative upper limit to $\eta$ from $Y_p \leq 0.248$ and $\DH \geq
1.5\times10^{-5}$.  The plot (a) is for the c.c geometry: the thick curves
are  for $f_r =1/\sqrt{2}$ (solid), 1/2 (dashed), $1/2\sqrt{2}$ (dot-dashed),
and the thin curves are for $f_r = 1/4$ (solid) and $f_r = 1/8$ (dashed).
The plot (b) is for the s.s. geometry: the thick curves are for $f_r = 1/4$
(solid), 1/8 (dashed), 1/16 (dot-dashed), and the thin curves are for
$f_r = 1/32$ (solid), 1/64 (dashed), and 1/128 (dot-dashed).
The allowed region is to the left of each curve.
}
\label{etafund}
\bigskip
\end{figure}

It turns out that all our IBBN models which satisfy Eq.~(\ref{Ypcons}) satisfy
Eq.~(\ref{DHism}) also.
In Fig.~\ref{etafund} we have plotted the contour~(\ref{Ypcons})
from all the models considered here. In SBBN the
constraint~(\ref{Ypcons}) gives an upper limit to $\eta$, $\et \leq 6.3$.
We see that, e.g., the centrally condensed IBBN models with $f_r = 1/8$
raise this upper limit to
\beq
   \et \leq 19, \mbox{\ \ or\ \ } \Omega_bh^2 \leq 0.07.
\label{maximus}
\eeq
Similar results were obtained for
the spherical shell geometry with $f_r = 1/32$
or 1/64; reaching this upper limit requires the distance scale to be close to
optimal in order to maximize the effect on $\eta$.

While IBBN raises the upper limit to $\eta$ from $\UHe$\ and D/H by a factor of
2 to 3, upper limits from $\ZLi$\ are raised at most by a factor of 1.4, and,
if we choose a very tight $\ZLi$\ limit, not at all.
Thomas \etal \cite{TSOMMF} used $\ZLiH < 1.4\times10^{-10}$, which
gives them an SBBN upper limit $\et \leq 3.1$, and this limit was not relaxed
at all by IBBN.  We confirm that none of our IBBN models raises the upper limit
to $\eta$ from this constraint.  However, their upper limit for $\ZLi$ allows
essentially no depletion at all.

\begin{figure}[tbh]
\vspace*{-0.3cm}
\epsfysize=8.2cm
\epsffile{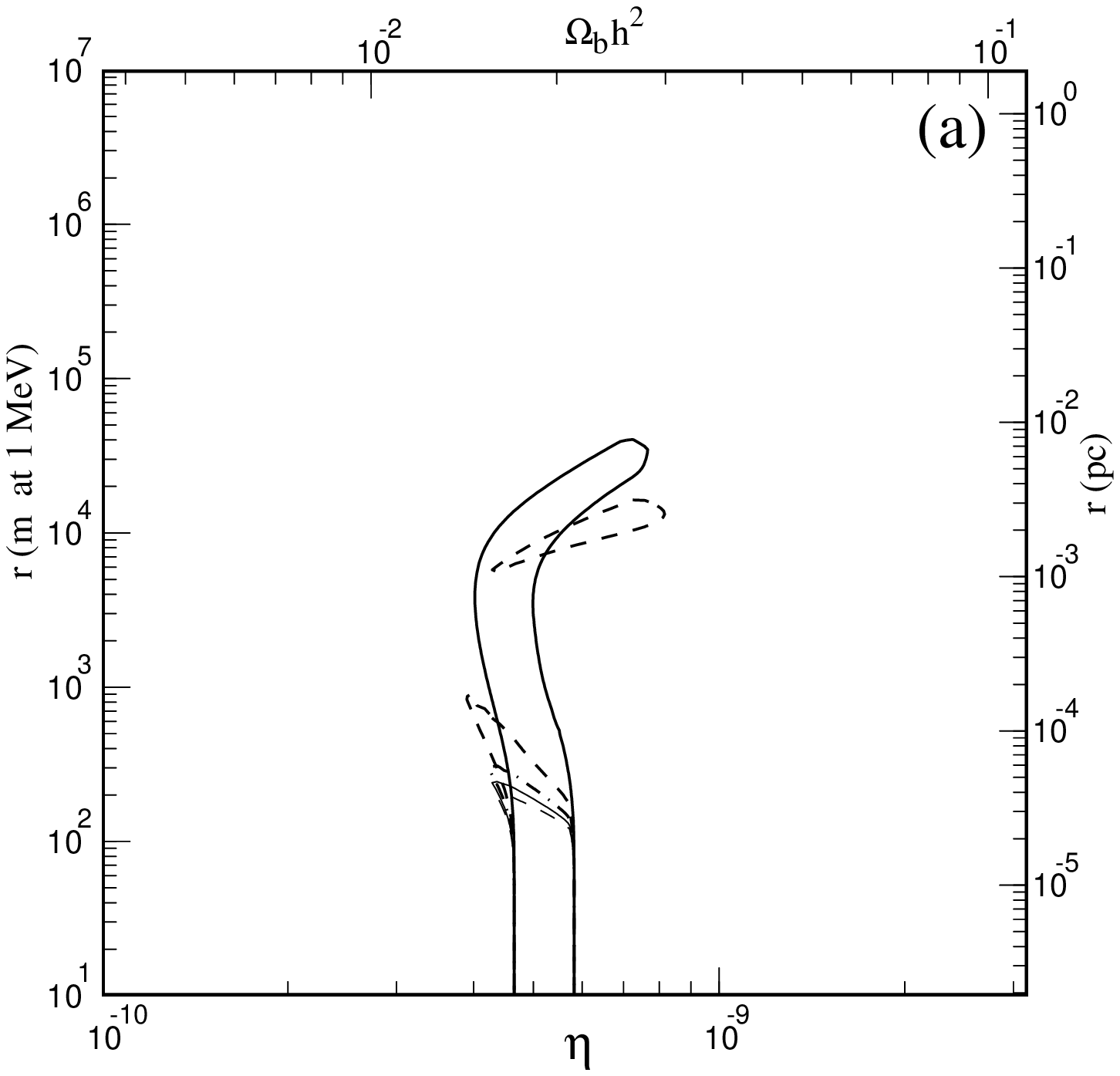}
\vspace*{-0.1cm}
\epsfysize=8.2cm
\epsffile{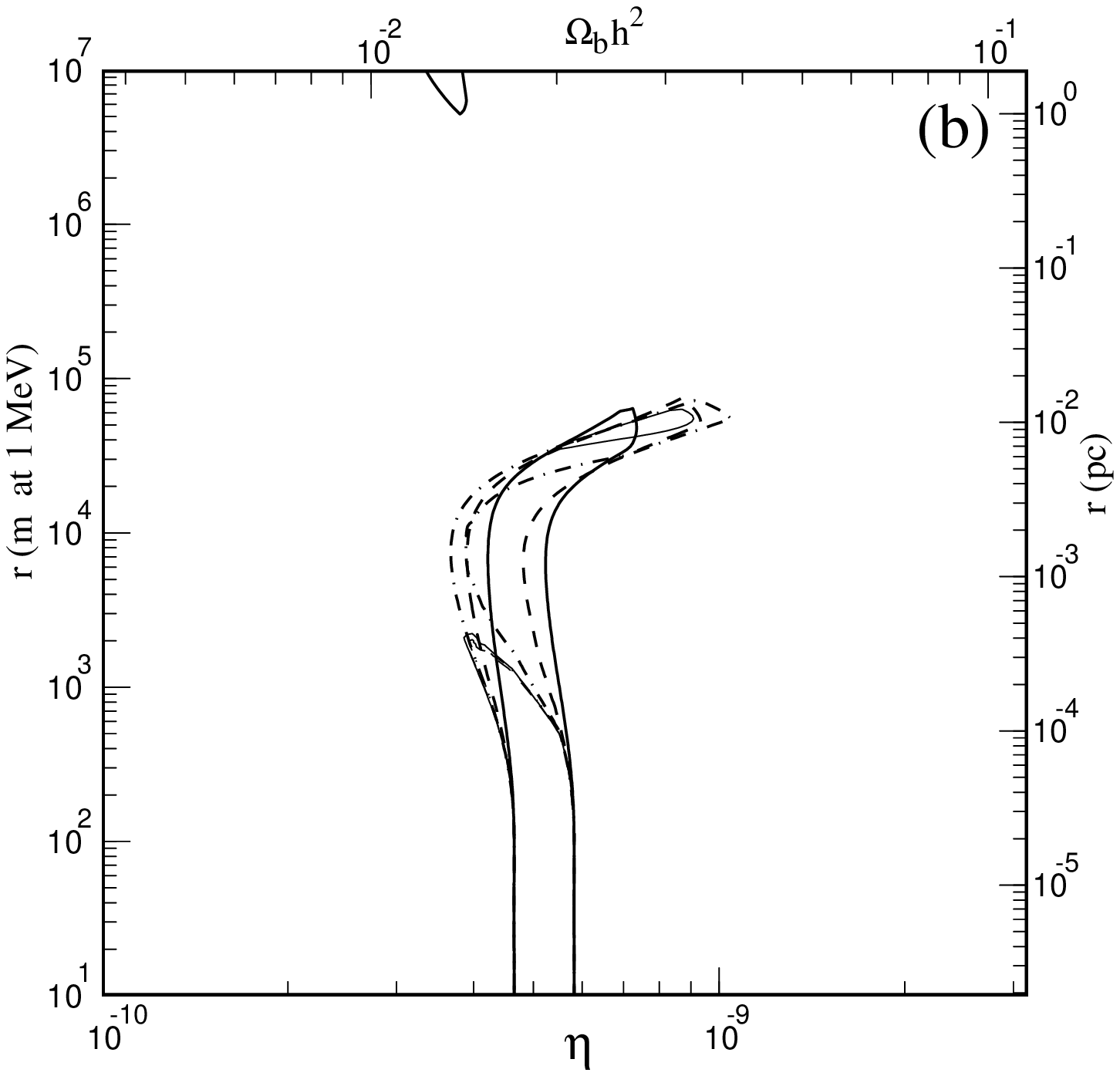}
\caption[a]{\protect
   The regions in the $(r,\eta)$-plane allowed by
D/H = $3.4\pm0.6\times10^{-5}$,
$Y_p \leq 0.248$, and $\log_{10}\ZLiH = -9.45\pm0.20$.  Plot (a) is for the
c.c.\ geometry and plot (b) is for the s.s.\ geometry.
The meaning of the
different line styles is the same as in Fig.~\ref{etafund}.
}
\label{higheta}
\bigskip
\end{figure}

As our second set we take the case for a high $\eta$ based on the high-$z$
deuterium value of Burles and Tytler \cite{BT98}.
We use the 2-$\sigma$ range 
\beq
   \DH = 3.4\pm0.6\times10^{-5}
\eeq
as our constraint.
For $\UHe$\ we continue to use the Izotov, Thuan \cite{IT98}
upper limit $Y_p \leq 0.248$ and for
$\ZLi$\ we use the Pinsonneault~\etal\ \cite{PSW}
range
\beq
   \log_{10}(\ZLi/{\rm H})_p = -9.45\pm0.20,
\eeq
further relaxed by the theoretical uncertainties as discussed in
Sec.~\ref{sec:com}.
The results for this set are displayed in Fig.~\ref{higheta}.

In SBBN these constraints lead to a baryon density in the narrow range
$\et = 4.6\mbox{--}5.8$.  In IBBN the allowed range is
\beq
   \et = 3.9\mbox{--}8.2
\eeq
for the centrally condensed geometry and
\beq
   \et = 3.7\mbox{--}10.5
\eeq
for the spherical shell geometry.

\begin{figure}[tbh]
\vspace*{-0.3cm}
\epsfysize=8.2cm
\epsffile{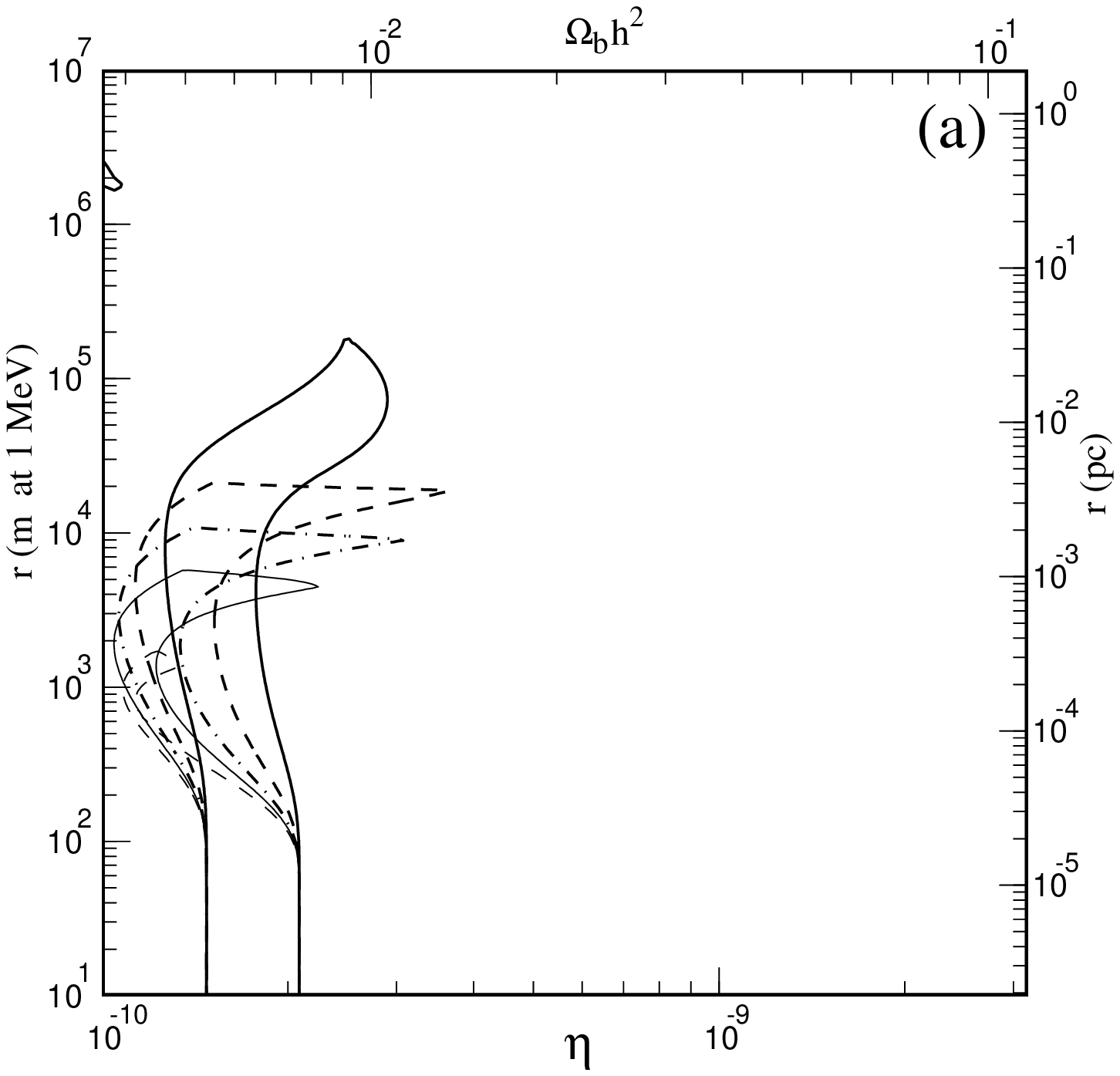}
\vspace*{-0.1cm}
\epsfysize=8.2cm
\epsffile{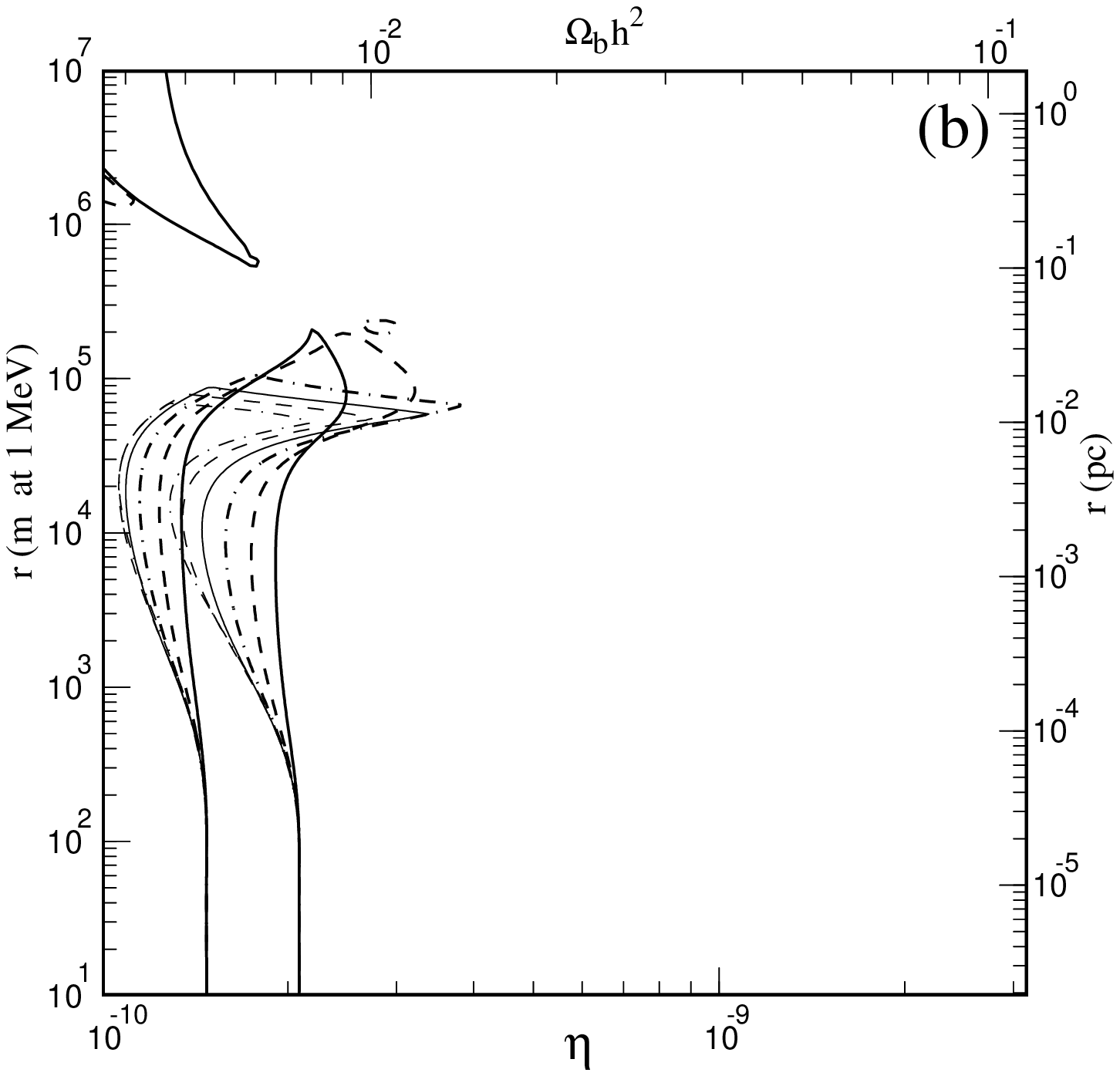}
\caption[a]{\protect
The case for low $\eta$.  This figure is similar to Fig.~\ref{higheta},
but the constraints used are $Y_p = 0.230\pm0.006$, $\DH \leq
2.5\times10^{-4}$, and $\log_{10}\ZLiH \leq -9.55$.
}
\label{loweta}
\bigskip
\end{figure}

In our third set we consider the case for low $\eta$ in SBBN \cite{fkot}.
(See Fig.~\ref{loweta}).  The 2-$\sigma$ OSS97 limits for $Y_p$
\beq
    0.224 \leq Y_p \leq 0.236,
\eeq
correspond to $\ZLi$\ near the Spite plateau and a large primordial D.
Hence we here use a conservative upper limit to D,
\beq
    \DH \leq 2.5\times10^{-4},
\eeq
and the Vauclair, Charbonnel \cite{VC98} upper limit for $\ZLi$,
\beq
    \log_{10} \ZLi/{\rm H} \leq -9.55.
\eeq
The results for this set are given in Fig.~\ref{loweta}. The SBBN range
is  $\et = 1.5\mbox{--}2.1$ (lower limit from D/H, upper limit from $Y_p$).
The IBBN upper limits are higher:
\bea
    \et & \leq 3.6 \mbox{\ \ (c.c.,\ }f_r = 1/2) \\
    \et & \leq 3.8 \mbox{\ \ (s.s.,\ }f_r = 1/16).
\eea

\begin{figure}[tbh]
\vspace*{-0.3cm}
\epsfysize=8.2cm
\epsffile{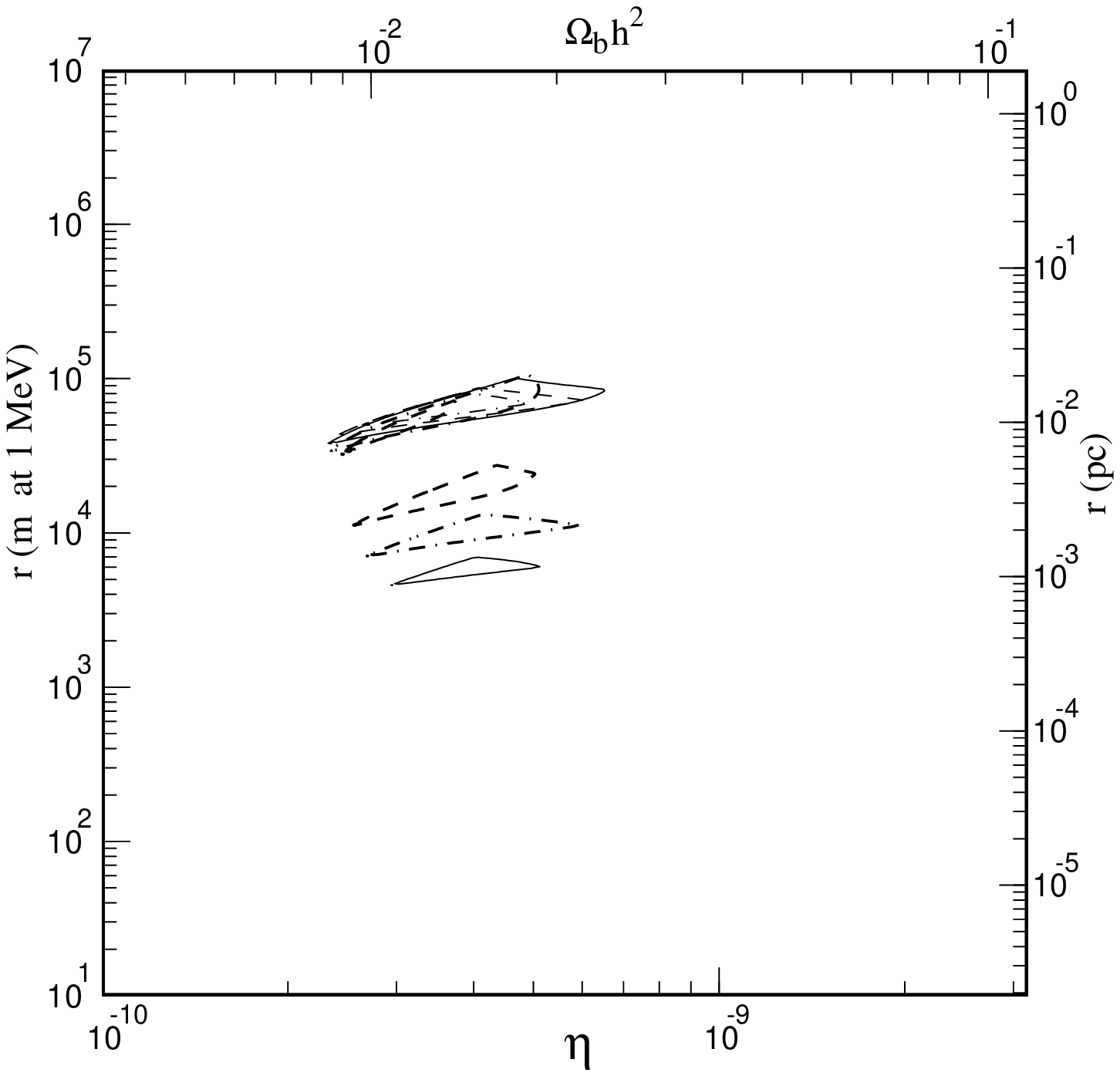}
\caption[a]{\protect
Alleviating the BBN ``crisis''.  This figure is similar to
Figs.~\ref{higheta} and \ref{loweta}, but the constraints used are
$Y_p \leq 0.238$, $\DH \leq 10^{-4}$ and $\ZLiH \leq 10^{-9.25}$.
These constraints are incompatible with each other in SBBN, but are
compatible in IBBN with the optimal distance scale.  The allowed regions for
the two geometries are shown in the same plot.  The ones for the s.s.\ geometry
are all at the same distance scale and lie on top of each other.  For the
c.c.\ geometry we get allowed regions for three of the considered volume
fractions, and they lie below the s.s.\ regions.
}
\label{solvecrisis}
\bigskip
\end{figure}

We finally demonstrate that IBBN can alleviate the tension between low $\UHe$
and low D.  If we use the constraints
\bea
    Y_p & \leq 0.238,  & \mbox{\ \ (SBBN\ } \et \leq 2.4) \\
    \DH & \leq 10^{-4}, & \mbox{\ \ (SBBN\ } \et \geq 2.6) \\
    \log_{10} \ZLi/{\rm H} & \leq -9.25, & \mbox{\ \ (SBBN\ } \et \leq 8.3)
\eea
no value of $\eta$ is allowed in SBBN (the ``crisis''). However, as is shown
in Fig.~\ref{solvecrisis}, some IBBN models satisfy these constraints, with
$ 2.6\leq \et \leq 6.0 $ (\cc) or $ 2.3 \leq \et \leq 6.5 $ (\ssh), in a narrow
(about a factor of two) range of the inhomogenity distance scale $r$.  This is
the ``optimum'' distance scale, which for these values of $\eta$ varies between
5 km and 30 km (at 1 MeV) for the centrally condensed geometry.
Similar solutions were
found with $r$ about 70 km for the spherical shell geometry, proving
that the result essentially depends only on the scale and is robust against
using different geometries.

\section{Conclusions}

In conclusion, we have studied the possibility of inhomogenous
nucleosynthesis on the basis of the new observational situation,
paying attention to the particular mechanisms capable of producing
the inhomogeneities in the very early universe.

First we studied the typical foam like inhomogeneity generated
during the electroweak phase transition, which we modelled by using
spherical symmetry with thin shells of high density regions. We find that
the scale from the EW transition tends to be too small to cause large
deviations from SBBN predictions; that is, the bound on $\eta$ is not
significantly changed.  However, the effects on theoretical yields
{\em can be} of equal size or larger than some of the more detailed
corrections recently included into the SBBN computations. Due to the
genericity of the EW-inhomogeneities these corrections can be claimed
to set the scale of accuracy achievable in SBBN computations.

Second we considered the full parameter space of the IBBN models in both
centrally condenced (QCD-type) and spherical shell (EW-type) geometries.

To answer the first question posed at the end of Sec.~\ref{sec:ibbn}: IBBN
models can satisfy the observational constraints equally well, and for some
small region of the parameter space, even better than SBBN.
For inhomogeneities with distance scales near the ``optimum''
scale $\ropt$, where the inhomogeneity
effects are maximized, this agreement is obtained for a larger baryon density
than in SBBN; precise values depend {\em intrinsically} on the observational
constraints, but the upper limit to $\eta$ from the upper limit to $\UHe$
and from the lower limit to $\DH$ may be raised by a factor of 2--3,
whereas upper limits set by the $\ZLiH$ data are raised less, at most by a
factor of 1.4.  However, it is not possible ever to get $\eta$ large
enough to make $\Omega_b=1$. For smaller scales the agreement is obtained
for similar or slightly smaller values of $\eta$ as in SBBN.

Regarding the second question, this optimum distance scale is not only larger
than the EWPT horizon, but it is
also several orders of magnitude larger than the
QCD transition distance scale
favoured by QCD
lattice calculations of the surface tension and the latent heat.  However,
the uncertainty in these values is as large as the values themselves so that
a much smaller latent heat, leading to a larger distance scale, is still
allowed; thus we cannot presently rule out the possibility of reaching the
optimum inhomogeneity distance scale in the QCD transition.

There is a region of parameter space, where the tension between $\UHe$ and D/H
is alleviated compared to SBBN. This takes place if the inhomogeneity
distance
scale is close to $\ropt$. The effect is however rather small, and for a low
deuterium, say $\DH \leq 5\times10^{-5}$, we cannot accommodate less helium
than
$Y_p = 0.240$, so IBBN cannot present itself as a solution to a dicothomy in
observations. Since we also pointed out that the present Kamiokande result
rules
out the simplest particle physics solution to possible tension in SBBN, the
conclusion,
that the problems are probably associated with the observations, is bolstered.

\section*{Acknowledgements}

We thank
the Center for Scientific Computing (Finland) for computational resources.

\end{document}